\documentclass[paper,twocolumn,twoside]{geophysics} 
\usepackage{graphicx}
\usepackage{subfigure}
\usepackage{subfiles}
\usepackage{algorithm2e}
\usepackage{amsmath,amsfonts,amssymb, mathtools}
\usepackage{subfig}
\usepackage{enumitem}

\begin{document}
\parindent=8pt

\newcommand{\rs}[1]{\mathstrut\mbox{\scriptsize\rm #1}}
\newcommand{\rr}[1]{\mbox{\rm #1}}

\onecolumn 

\begin{enumerate}

\item[\textbf{Citation}]{O.J. Aribido, G. AlRegib, and Y. Alaudah, ``Self-Supervised Delineation of Geological Structures using Orthogonal Latent Space Projection," in Geophysics, vol. 86, no. 6, accepted Jul. 29 2021}


\item[\textbf{Review}]{Accepted for publication: 29 July 2021}

\item[\textbf{Codes}] {https://github.com/olivesgatech/Latent-Factorization}

\item[\textbf{Bibtex}]{\fontfamily{lmtt}\selectfont {@article \{aribido2021self, \\
author = {Aribido Oluwaseun Joseph and Ghassan AlRegib and Yazeed Alaudah}, \\
title = {Self-Supervised Delineation of Geological Structures using Orthogonal \\ Latent Space Projection}, \\
journal = {GEOPHYSICS}, \\
volume = {86}, \\ 
number = {6}, \\
year = {2021}, \\
\}
}}


\item[\textbf{Copyright}]{\textcopyright 2021 \textbf{Geophysics}. A revised version of this manuscript has been accepted to Geophysics and is awaiting production. The copyrights for the accepted manuscript belong strictly to the Society for Exploration Geophysicists (SEG). This document
may strictly be used only for educational and other non-commercial purposes only. The full citation to the accepted
manuscript will be made available once the DOI has been published.}

\item[\textbf{Contact}]{mailto: oja@gatech.edu  \\ 
mailto: alregib@gatech.edu, Website: https://ghassanalregib.info  \\
mailto: yalaudah@gmail.com\\
}

\thispagestyle{empty}
\newpage
\clearpage
\setcounter{page}{1}

\twocolumn

\end{enumerate}

\title{Self-Supervised Delineation of Geological Structures using Orthogonal Latent Space Projection}

\address{
\footnotemark[1] Center for Energy and Geo Processing (CeGP), 
GA 30332-0250, U.S.A.,
\footnotemark[2] King Fahd University of Petroleum and Engineering (KFUPM)}

\author{Oluwaseun Joseph Aribido\footnotemark[1], Ghassan AlRegib\footnotemark[1], Yazeed Alaudah\footnotemark[2]}

\renewcommand{\thefootnote}{\fnsymbol{footnote}} 

\footer{Example}
\lefthead{Aribido, AlRegib \& Alaudah}
\righthead{Delineation of Geological Structures}


\maketitle

\begin{abstract}
    \hspace{8pt}We developed two machine learning frameworks that could assist in automated litho-stratigraphic interpretation of seismic volumes without any manual hand labeling from an experienced seismic interpreter.  The first framework is an unsupervised hierarchical clustering model to divide seismic images from a volume into certain number of clusters determined by the algorithm. The clustering framework uses a combination of density and hierarchical techniques to determine the size and homogeneity of the clusters.  The second framework consists of a self-supervised deep learning framework to label regions of geological interest in seismic images. It projects  the  latent-space  of  an  encoder-decoder architecture  unto  two  orthogonal  subspaces,  from  which  it  learns  to delineate  regions  of interest in the seismic images. To demonstrate an application of both frameworks, a seismic volume was clustered into various contiguous clusters, from which four clusters were selected based on distinct seismic patterns: horizons, faults, salt domes and chaotic structures. Images from the selected clusters are used to train the encoder-decoder network. The output of the encoder-decoder network is a probability map of the possibility an amplitude reflection event belongs to an interesting geological structure. The structures are delineated using the probability map. The delineated images are further used to post-train a segmentation model to extend our results to full-vertical sections. The results on vertical sections show that we can factorize a  seismic volume into its corresponding structural components.  Lastly, we showed that our deep learning framework could be modeled as an attribute extractor and we compared our attribute result with various existing attributes in literature and demonstrate competitive performance with them.
\end{abstract}

\section{Introduction}
\label{introduction}


\hspace{8pt}Seismic interpretation requires detailed understanding of seismic acquisition, processing, and data models to infer geological meaning. The process of seismic data preprocessing and migration involves geometric transformation and analysis to produce an accurate image of Earth's subsurface. Prestack and poststack preprocessing can introduce artifacts or coherent noise into seismic data, leading to false-positive identification of seismic structures during interpretation. Hence, consistent accurate interpretation of seismic data requires many years of experience. 

Despite improvement in the quality of 3-D migrated seismic data, thorough interpretation of certain geologic elements remains subjective. Subjective interpretation occurs due to the presence of many `valid' interpretations or weak amplitude reflections. One cause of weak amplitude reflection is the absence of bottom simulating reflectors in the subsurface during acquisition \cite[]{bedle2019seismic}. In addition, interpreted seismic data are considered intellectual property by energy industries. Consequently, publicly annotated data are scarce. All these challenges necessitate the need for a model-based framework that is objective to defined constraints, requires little or no human-assisted label, and powerful enough to learn deep-diverse patterns in seismic data.
Recent deep learning success stories have further motivated research for automated seismic interpretation using machine learning. But the limited label problem is a challenge to training deep learning models imported from computer vision applications where labels ramp into millions in number. Consequently, deep learning models designed for seismic applications must be trained with limited data awareness. 

\hspace{8pt}To address these challenges, we propose a \emph{self-supervised} learning framework that does not require any labels from interpreters. Rather, the model is trained on the physics of seismic patterns, from which homogeneous patterns are separated out using constraints. 
Various computer-assisted frameworks have been proposed in the literature, of which we make two broad categories: attributes-based interpretation and machine learning-based interpretation. Seismic attributes-based methods rely on mathematical computations to identify distinctive patterns of seismic amplitudes. These patterns are mapped to a database of previous successful patterns for successful interpretation \cite[]{chopra2005seismic}. Because the computation of geometric attributes can be automated, many seismic attributes have been introduced to the geoscience community \cite[]{taner1994seismic, barnes1992calculation, chen1997seismic, shafiq2015detection, shafiq2017salt, shafiq2018role, shafiq2018novel}. However, seismic attributes are usually designed to identify specific patterns of interest to an interpreter. Hence, patterns that deviate from the specific target are unidentified. This implies a suite of complementary attributes must be selected by an interpreter interested in identifying important events. Our proposed model, however, learns patterns without previous specification bias. In addition, it contains millions of parameters which is powerful enough to learn very complex patterns that would be missed by simpler algorithms such as attributes.

\hspace{8pt}Early adoption of machine learning models in seismic research began with supervised methods in which the model has access to labeled data. \cite[]{di2018patch, wu2018convolutional, xiong2018seismic, araya2017automated, dramsch2018deep}.  Several notable works have also attempted to overcome the effect of limited annotated data by employing semi-supervised and weakly supervised techniques \cite[]{alaudah2016weakly,  alaudah2018structure, di2019semi,babakhin2019semi, alfarraj2019semi, wu2019faultseg3d, liu20193d, di2018deep}. In these frameworks, there is less dependence on fully labeled data. In semi-supervised frameworks, for instance, researchers use fewer annotations augmented with pseudo-labels. Weakly supervised learning models use weaker labels like image labels only \cite[]{alaudah2018learning}. Weak labels are easier to generate in numerous quantities compared to pixel-level annotations required in supervised frameworks. 

\hspace{8pt}In contrast, our method does not require any annotated labels from interpreters. This ranks it at the same ease of use as attributes, with the exception that we applied a more powerful learnable model. Another relevant body of literature explores unsupervised learning works which also addresses the limited data problem. These frameworks are mostly based on K-Means clustering and Konohen's self-organizing maps (SOMs) \cite[]{barnes2002investigation}. The basic workflow includes extracting attributes from seismic data and using a dimension-reduction algorithm, mostly principal component analysis (PCA), to identify the most important features. The principal features are then clustered to a specified number of centroids. Although these methods have produced great results, the ground-truth centroids of the attributes are unknown; and manually initializing centroids does not converge to ground-truth centroids. Secondly, PCA leads to information loss and the distance metric used in clustering algorithms is usually Euclidean based, which affects the cluster accuracy. Our proposed clustering framework does not require the number of centroids to be predefined, rather our algorithm explores multi-scaled, directional spectral information in the images to extract high-dimensional coefficients before clustering them using a custom-distance metric. 

\hspace{8pt}Lastly, we explore other relevant literature based on their target applications. For instance, detection of faults \cite[]{araya2017automated, di2019improving, di2019semi, shafiq2018novel,wu2019faultseg3d, xiong2018seismic}, delineation of salt bodies, \cite[]{di2018multi, di2018deep, shafiq2015detection}, classification of facies \cite[]{liu20193d, dramsch2018deep, qian2017seismic, alaudah2019machine, alaudah2019facies}, prediction of rock lithology from well logs \cite[]{alfarraj2019semi, das2018convolutional, das2019effect} and segmentation of seismic layers are few areas of relevant applications of deep learning to seismic. In these literature, various depth of labels are used to train the network. Our proposed method eliminates this challenge by learning image labels using an unsupervised framework.  \cite{dubrovina2019composite} propose using the latent space of an encoder-decoder architecture to split and rearrange various parts of a 3D object. The latent space is projected to summable orthogonal subspaces. Each orthogonal subspace of the latent variable retains low-level features of various parts of the 3D object. However, the dataset includes ground-truth labels of 3D parts, our method does not include any ground-truths for partitioned parts. \cite{li2019orthogonal} add an orthogonal penalty to latent variables and show that by using SOMs on the orthogonal features, more separability on pixel-space features is realized. The methodology presented in \cite{li2019orthogonal} is similar to ours in applying orthogonality to latent variables. In our self-supervised framework, we introduced projection matrices to factorize input images; hence, eliminating the need for SOM on the features.

\hspace{8pt}In this work, we use the F3 block as our dataset. The F3 block is an offshore block in the Dutch sector of the North Sea. The dataset is preprocessed using a dip-median filter to remove random noise and  to enhance the edges of the seismic reflections. Seismic traces in the volume are clipped above and below 4.0 times the standard deviation of the volume. All amplitude values are normalized to the range [-1, 1]. The volume is split into 120x120 overlapping patches. We propose a new hierarchical clustering model to group these patches into $K$ classes. Four classes out of all classes are passed to a deep encoder-decoder model, augmented with two discriminators. The latent space of the encoder-decoder model is projected unto two learned subspaces. We added constraints to guide the factorization of each input patch into two images. Each synthesized image corresponds to a learned subspace. 
The subspaces are further constrained to be orthogonal. Though we use discriminators, our method is self-supervised because the adversarial part of our model is used in a multi-task setting aside from learning the factorized latent space, which is done unsupervised. Lastly, we conclude by evaluating our proposed method against related research. We further show that our deep encoder-decoder model learns attributes that are qualitatively better than traditional attributes for structural delineation.

\section{Hierarchical Clustering Framework}

\hspace{8pt}We propose a hierarchical clustering framework to cluster images into $K < 20$ clusters. Where $K$ is the number of clusters.  The volume is subdivided into subset blocks. Each subset block contains 15 vertical sections in the inline or crossline direction. We separate inline and crossline subset blocks. The clustering framework consists of several layers. The first layer is initialized using a density-based algorithm: density-based spatial clustering of applications with noise (DBSCAN) \cite[]{ester1996density}, to produce contiguous clusters. All other layers consist of hierarchical merges of clusters in proceeding layers. Three sections are extracted from each subset block taken at five intervals apart. A set of three sections at intervals of five apart, is called a category. Hence, each subset-block contains three categories.

\hspace{8pt}We do not randomize all the images collected from the volume to make them independent and identically distributed, to avoid losing domain correlation information between sections. However, within each category, we randomize the order of all images. Each category contains $270$ images and $165$ images for inline and crossline respectively. Inline categories are clustered separately from crossline categories due to differences in feature representation in both sets.

\subsection{Clustering with DBSCAN}
\begin{figure}[htb!]
    \centering
    \includegraphics[scale=0.34]{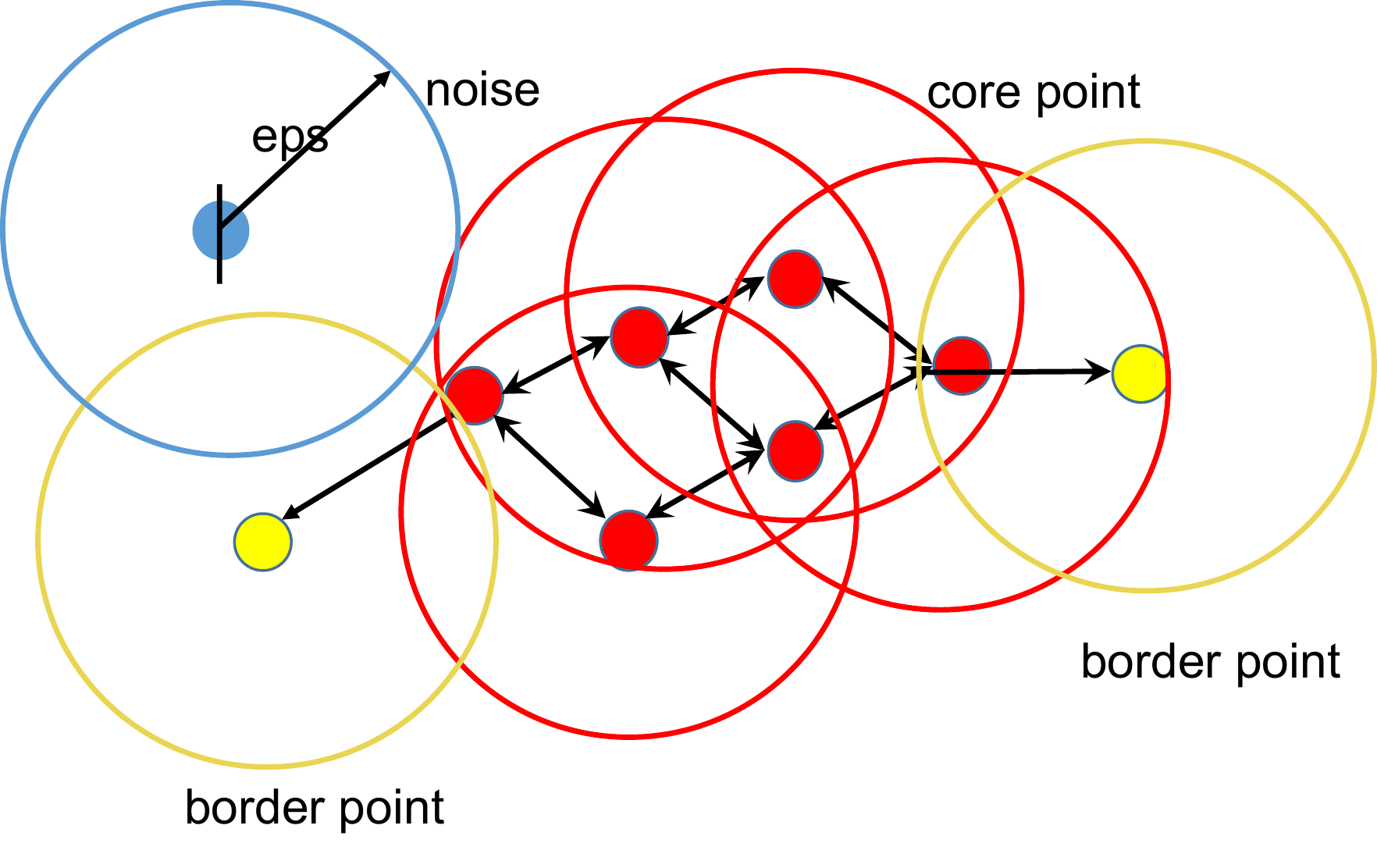}
    \caption{A toy example of DBSCAN algorithm,  $m=3$. The radius of the circles represent $\epsilon$ distance from each point. The red points are core points because they are within $m$ neighbors of each other. Yellows points are border points because they have less than $m$ points in their $\epsilon$ neighborhoods. The blue point is considered noise because it is not within $\epsilon$ distance to any other point.}
    \label{fig:dbscan}
\end{figure}

\hspace{8pt}DBSCAN uses density-based metrics for cluster discovery.  Two parameters are usually defined for DBSCAN: $m$ and $\epsilon$. $m$ is the minimum number of points that can form an independent cluster, while $\epsilon$ is the maximum distance between two core points. Given a group of points, DBSCAN determines whether a point is a core or border point. The former belongs to a cluster with at least $m$ minimum points within $\epsilon$ distance of each other. The latter is within $\epsilon$ distance from any core point but not within $\epsilon$ distance to $m$ points. At initialization, one cluster is randomly chosen from the dataset and neighboring points are checked to find whether they are core or border points. If $m$ points are core points, a cluster is established and other core points and border points $\epsilon$ distant away are added to that cluster. Points that are neither core nor border point, are considered noise. In Figure \ref{fig:dbscan}, we show a toy example of DBSCAN with $m=3$. The red points are core points because they have $m\ge3$ neighboring points at $\epsilon$ distance away. The yellow points are border points, because they are $\epsilon$ distant from less than $m$ points, but are in the neighborhood of a core point. All core points are considered density-reachable from other core points. The blue points are considered noise because they are not $\epsilon$ distant from any point in the group. 

\hspace{8pt}To specify a distance metric for DBSCAN between two seismic images, we used a similarity (SIM) metric proposed by \cite{alfarraj2016content}. The authors showed that SIM compares curvelet coefficients from two images. Curvelets coefficients \cite[]{starck2002curvelet} are multi-scaled decompositions of an image obtained by tiling the frequency domain with trapezoidal-shaped tiles \cite[]{alfarraj2016content}. Additionally, SIM outperforms many other similarity measures for texture images like S-SSIM \cite[]{wang2004image}, CW-SSIM \cite[]{gao2011cw}, and STSIM \cite[]{zujovic2013structural}. The tiled frequencies collected at various scales (here 32) of the image in various directions, make the algorithm a good edge detector. Hence, suitable for seismic images.  Singular value decomposition (SVD) is applied to the curvelet coefficients extracted from the images to select the best coefficients.  The reduced coefficients are then compared using Czenkanoski's similarity to obtain a score in the range [0, 1] \cite[]{alfarraj2016content}:

\begin{equation}
    SIM(I_1, I_2) = \frac{||v_1 - v_2 ||_1}{||v_1 + v_2||_1}.
\end{equation}

\noindent The $SIM(I_1, I_2)$ returns values from $[0, 1]$ for each image pair, where $I_1$ and $I_2$ are images and $v_1$ and $v_2$ are the SVD reduced curvelet coefficients. Hence, a $0$ implies perfectly similar, while $1$ denotes perfectly dissimilar.

Inline and crossline images are fed sequentially into the DBSCAN algorithm. By experimenting with several hyperparameter values, we set $\epsilon=0.10$, for inline images and $\epsilon=0.09$, for crosslines, while $m=3$ for both inline and crossline images. Thus, for any image occurring at any location within any section, if that image is not $\epsilon$ distant to any other image in the same category, it is marked as noise. We keep $\epsilon$ small to ensure thorough discrimination between clusters. The drawback of a small $\epsilon$ is that there will be  many $(\sim 30)$, small clusters per category. However, this remains solvable since a hierarchical algorithm is applied to merge contiguous clusters.

\subsubsection{Algorithm 1}

\begin{figure}[htb!]
    \centering
    \includegraphics[scale=0.4]{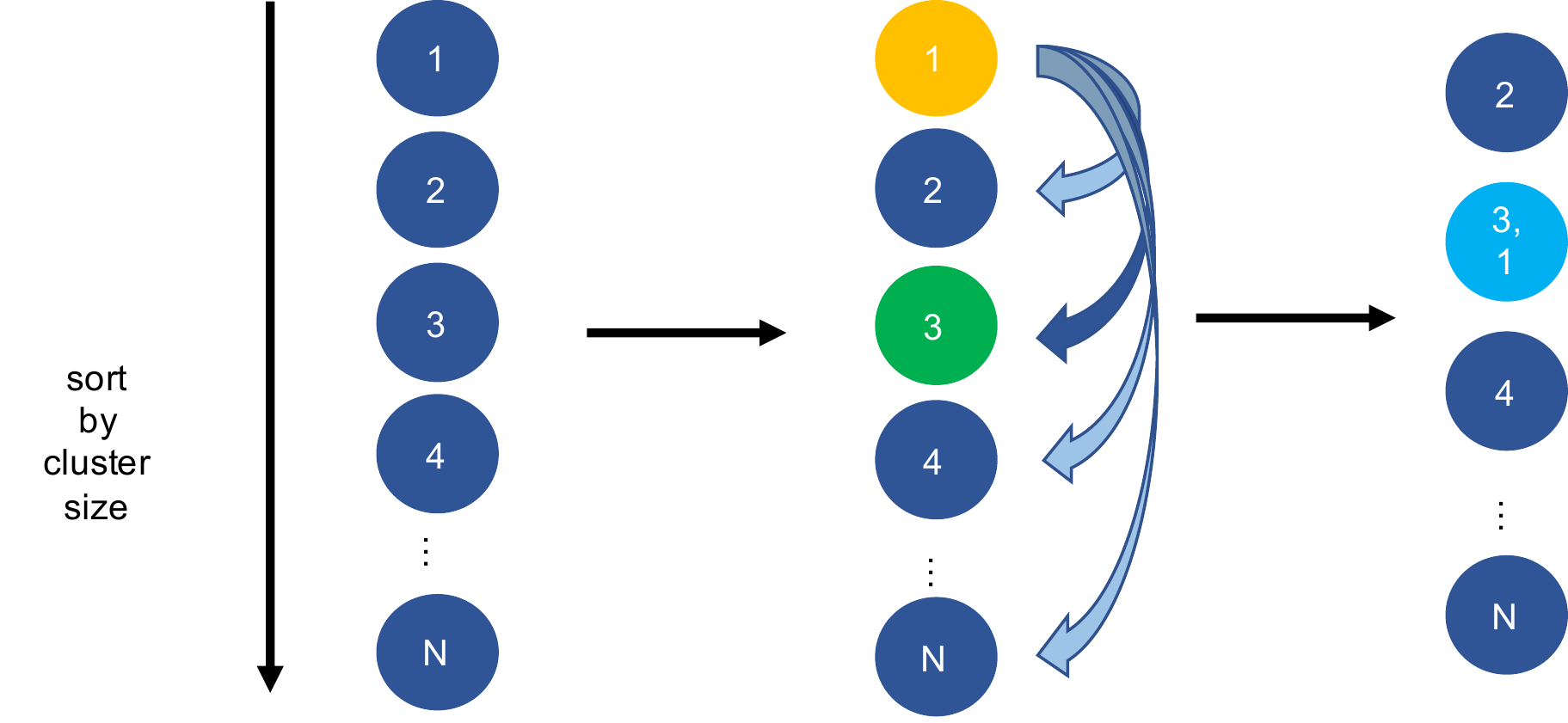}
    \caption{Algorithm 1 procedure for intra-category merging of clusters. The blue circles are clusters arranged in ascending order of magnitude. For one merge instance, we compare cluster 1 with all other clusters till we find the closest match, e.g., cluster 3. In which case, we merge clusters 1 and 3 and re-sort all clusters again.}
    \label{fig:algorithm1}
\end{figure}

\hspace{8pt}In the first step, we merge intra-category clusters. As shown in Figure \ref{fig:algorithm1}, we arrange the clusters in order of cluster size. Each circle represents a cluster. The blue column represent one category. We select a handful of images from the smallest cluster and compare these images to remaining clusters inside the category. The smallest cluster is then merged with the most similar cluster, in this case, cluster three. The algorithm is iterated until the number of clusters reduces from $\sim 30$ to a predefined small number (e.g., $8$).  

\subsubsection{Algorithm 2}
\begin{figure}[htb!]
    \centering
    \includegraphics[scale=0.40]{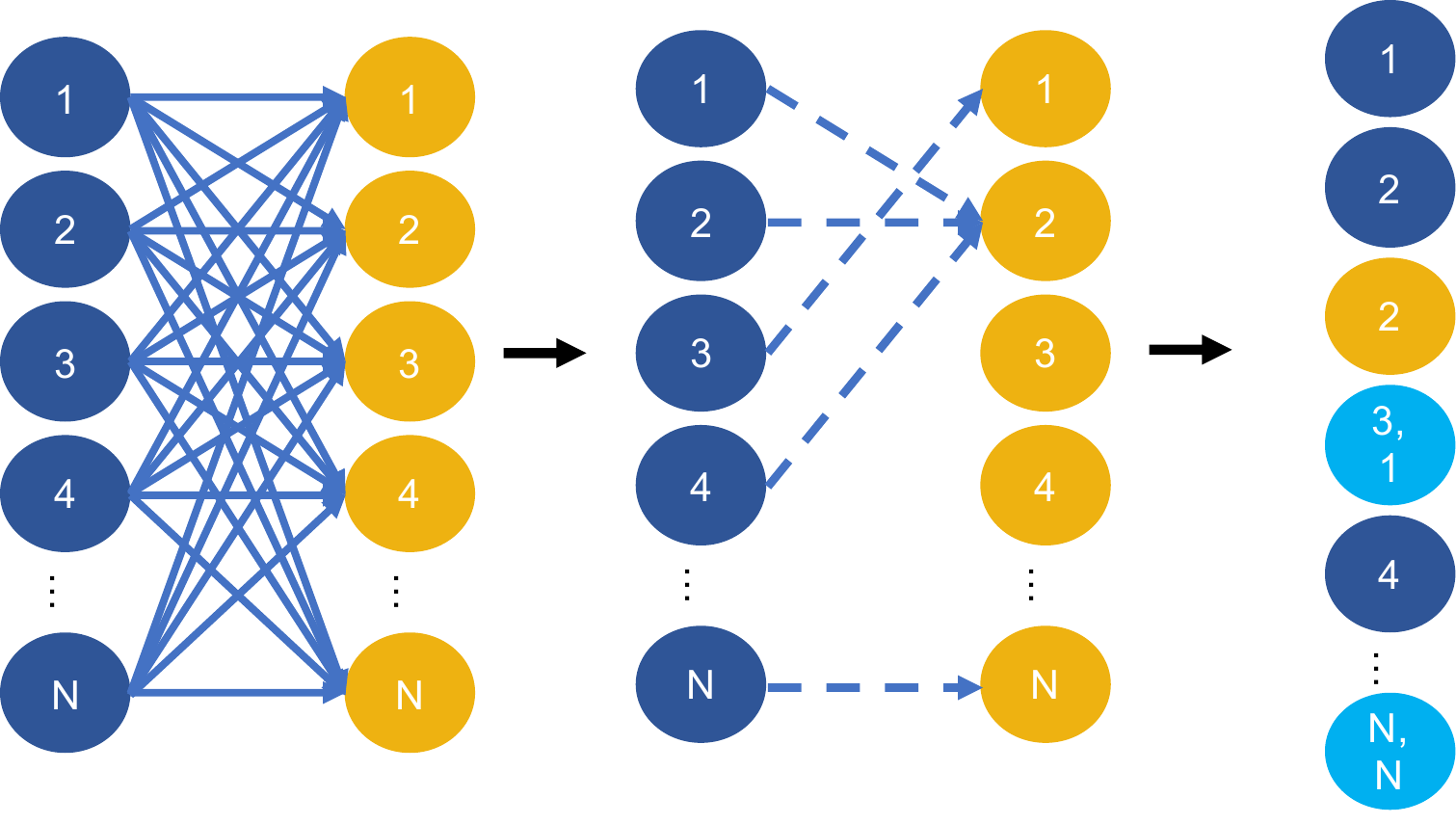}
    \caption{Algorithm 2 procedure for inter-category merging of clusters. The blue and yellow columns are two categories: $A$ and $B$, whose clusters are to be merged. We compare each cluster in $A$ with every other cluster in $B$ to find the closest match. In this illustration, clusters 1, 2 and 4 in $A$ all select cluster 2 in $B$ as their closest match. Cluster 3 in $A$ maps to cluster 1 in $B$. Assuming for all other clusters in $A$, we obtain a one-to-one mapping in close match to clusters in $B$, we only match clusters with one-to-one matching between $A$ and $B$. We then copy over clusters with many-to-one $(1,2 \mbox{ and } 4)$ mapping from $A$ to $B$.}
    \label{fig:algorithm2}
\end{figure}

\hspace{8pt}In the second step, we merge inter-category clusters. In  Figure \ref{fig:algorithm2}, the blue column is category $A$, the yellow column is category $B$. A few images are randomly sampled from each cluster in $A$ and compared with all clusters in $B$. All clusters in $A$ with a one-to-one mapping, in best proximity to clusters in $B$, are merged. For instance, clusters 1, 2, and 4 in $A$ all select cluster 2 in $B$ as their closest cluster (many-to-one mapping), we do not merge any of these clusters. However, we merge clusters 3-to-1 and N-to-N. After merging similar clusters, the remaining clusters in $B$ are transferred to $A$.

\begin{figure}[htb!]
    \centering
    \includegraphics[scale=0.40]{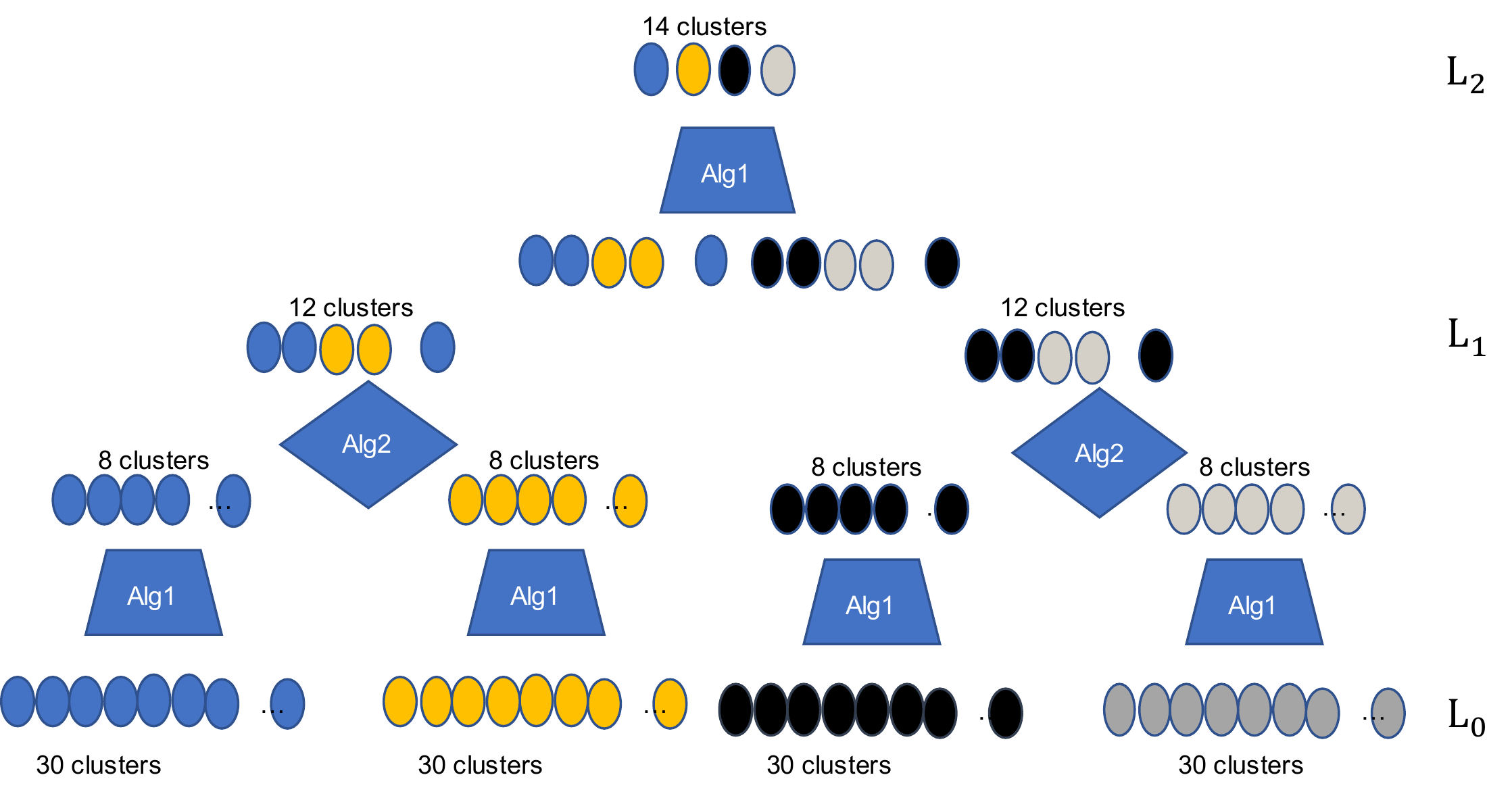}
    \caption{Hierarchical cluster merging. $L_0$ is the first layer and contains clusters from DBSCAN. Subsequent layers are obtained by iterating algorithm 1 and 2. $L_1$ is the second layer of clusters after the first iteration of both algorithms. $L_2$ is the third layer and so on.}
    \label{fig:hierarchy_merging}
\end{figure}

Figure \ref{fig:hierarchy_merging} illustrates how the hierarchical clustering module works. Both algorithms are applied alternatively on the data until the total number of categories are reduced to $8$ and the total number of clusters inside all categories are reduced to $14$. 
Lastly, we inspect all 14 clusters. The total number of clustered images was 6386. There are eight visually unique clusters, including noise. Hence, we combine repetitive clusters. For instance, dipping amplitudes and non-dipping amplitudes can be combined into the `horizons’ cluster. 

\begin{figure*}[!thb]
    \centering
    \includegraphics[scale=0.45]{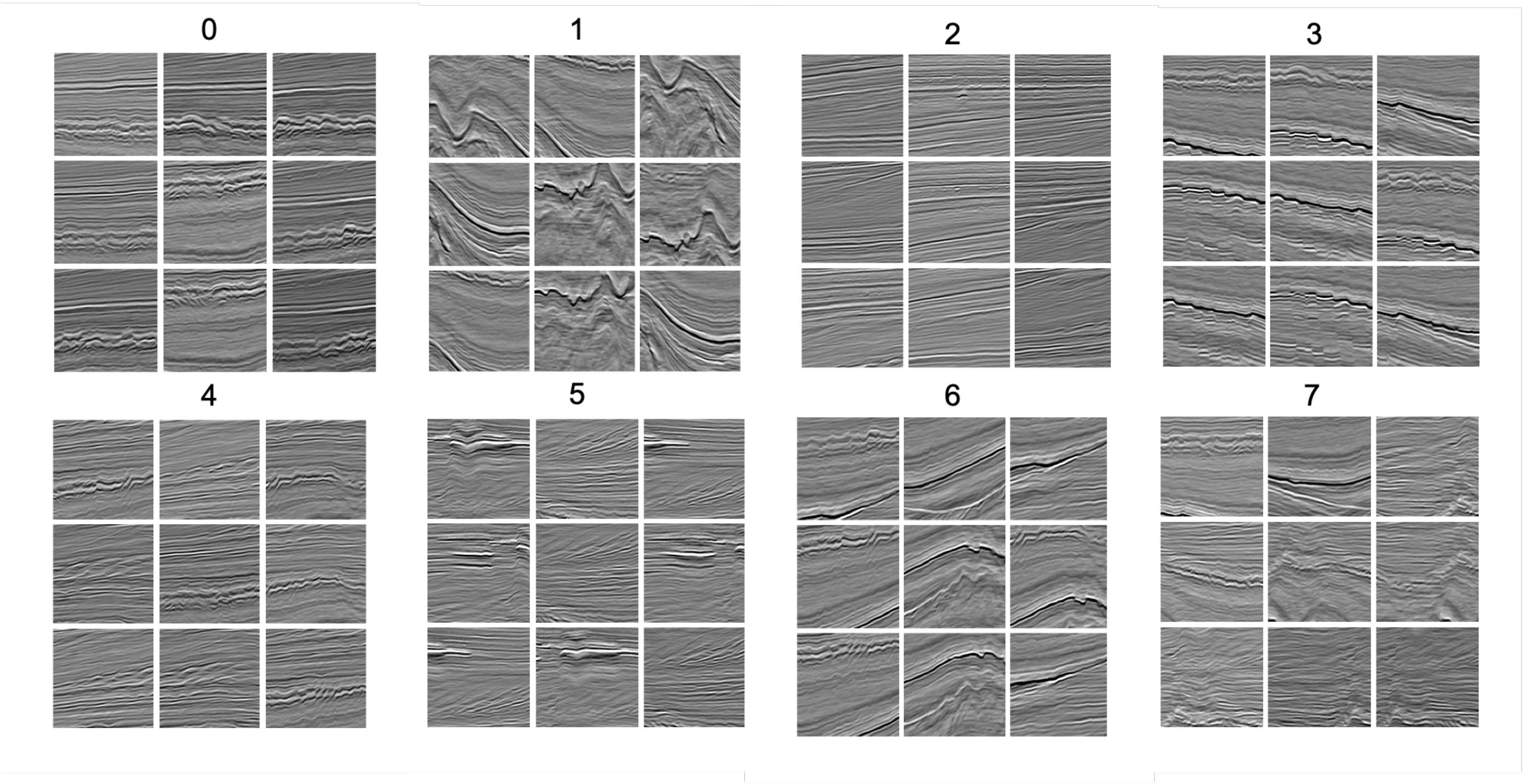}
    \caption{Result showing eight clusters at the final layer.}
    \label{fig:clustering_result}
\end{figure*}

\hspace{8pt}In Figure \ref{fig:clustering_result}, Cluster 0 is labeled as `chaotic' because the dominant geologic component highlighted by the clustering algorithm is the chaotic facies. Cluster 1 contains images from the salt dome region of the volume. In this cluster, it is interesting that although the patterns of the reflection amplitudes are diverse, our algorithm clusters them accurately. Cluster 2 identifies parallel reflectors, that we label as `horizons'. The horizons cluster contains the most number of images of all clustered images, and the clustering algorithm also attains the highest accuracy in identifying this cluster. 
\hspace{8pt} Cluster 3 captures faulted regions of the volume and are labeled as faults. Notice some discontinuities are labeled as faults. Cluster 4 captures another variant of the chaotic class with some non-conformity in seismic reflections. Cluster 5 contains images taken from the bright-spot region of the seismic volume. Interestingly, our algorithm differentiated this class of images from the horizon class in Cluster 2. Cluster 6 highlights images from the top of the salt dome and they were clustered into a different class by our algorithm because they did not fully represent structures captured inside the salt dome. Cluster 7 shows irregular seismic reflections. Two of them contain chaotic regions while others do not fit into any of the other classes. There are several classes like cluster 7 and 5 that would not be used in training our deep learning framework, but would be left for further study in future research. Figure \ref{fig:clustering_result} contains the output of our clustering framework.

\hspace{8pt}Four clusters were selected from Figure 5 for training the deep learning model. These clusters would be referred to as classes in future references. From the result in Figure 5 Horizons (class 2), faults (class 3), salt domes (class 1) and chaotic (class 0) labels are representative of the selected images. The `horizon' cluster is the largest class with 1258 images and the Chaotic class is the smallest class with 389 images. The Fault and the salt dome classes had 720 and 500 images, respectively. We discarded images clustered as noise. Hence, for balanced classes, we randomly select 389 images from each class, making a total of 1556 training images.

\section{Deep Learning Framework}
\label{lt_fact}

\begin{figure}[!thb]
    \centering
    \includegraphics[scale=0.46]{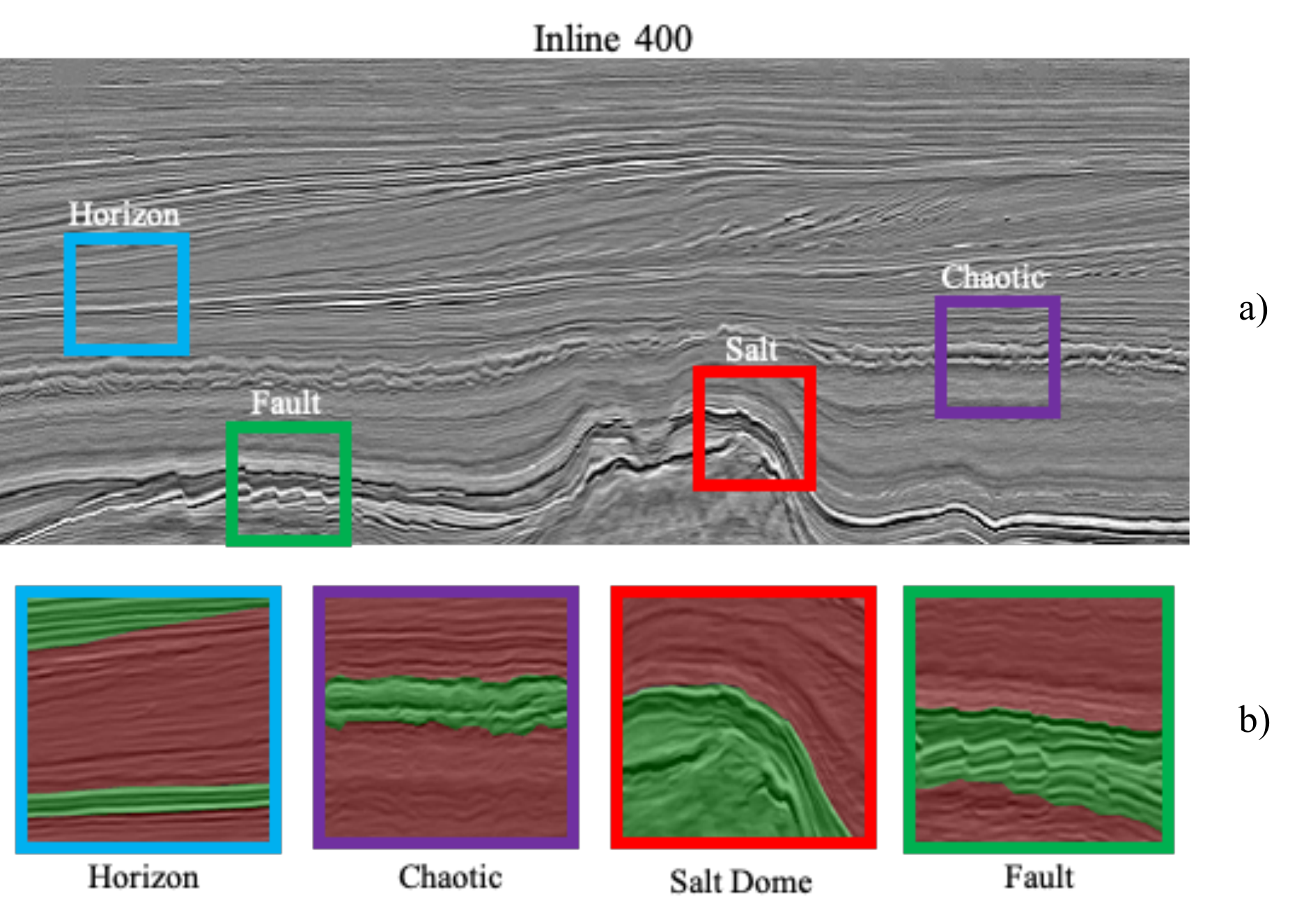}
    \caption{A sample section image from inline 400 (a) showing four samples of images clustered by our hierarchical model, (b) shows manually labeled structural components versus background. Structural components are annotated in green and the background in red.}
    \label{fig:sample_patches}
\end{figure}
\begin{figure*}
    \centering
    \includegraphics[scale=0.5]{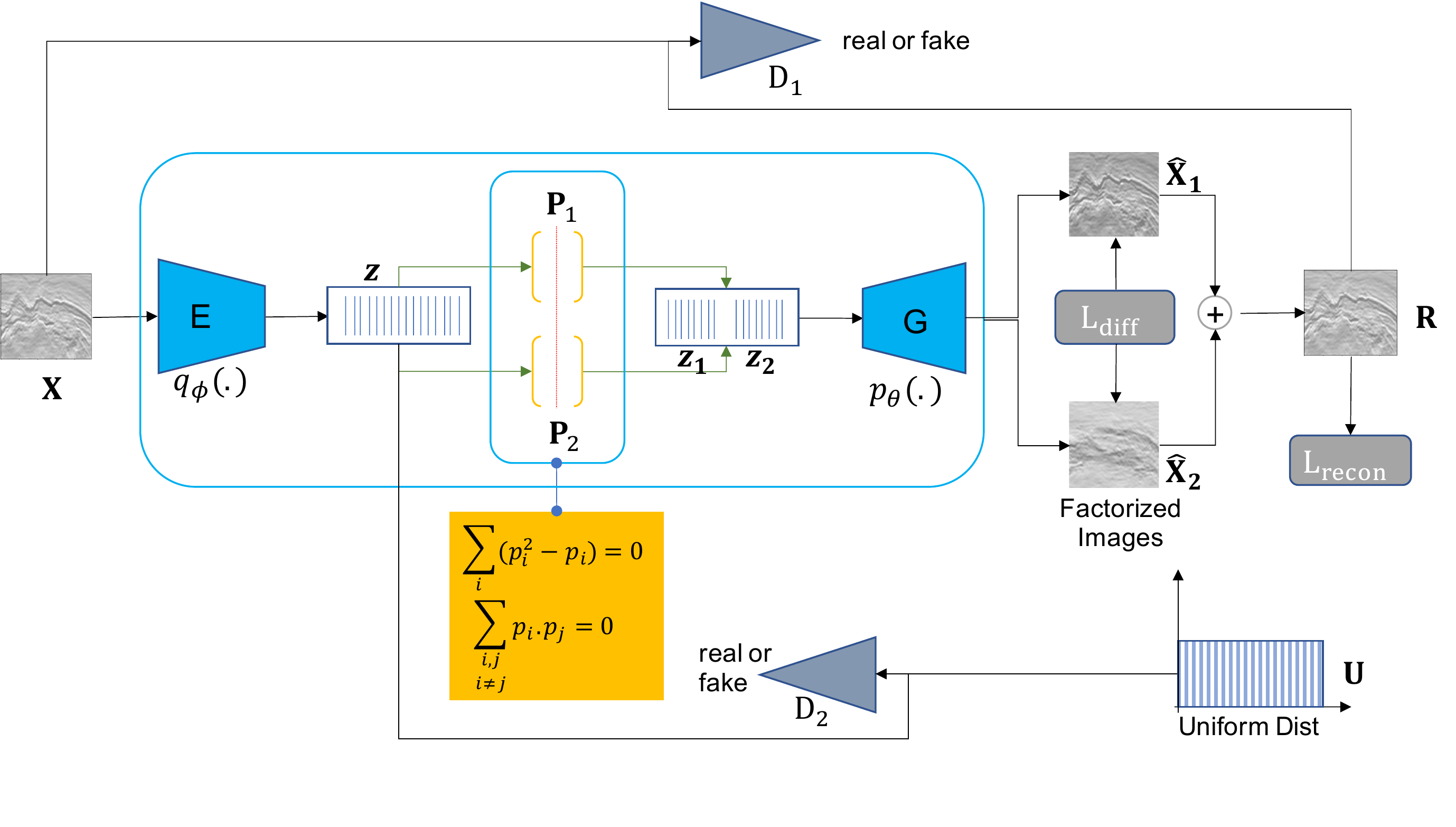}
    \caption{Diagram of proposed deep learning module. $\mathbf{X}$ is the set of all input images. $q_{\phi}(.)$ is the encoder. $\mathbf{z}$ is the latent code factorized into $\mathbf{z}_1, \mathbf{z}_2$. $p_{\theta}(.)$ is the decoder that outputs the factorized images: $\hat{\mathbf{X}}_1$ and $\hat{\mathbf{X}}_2$.} 
    \label{fig:model}
\end{figure*}

 
\hspace{8pt}We have a set of clustered seismic images in four classes, each with equal number of images. The assignment of image labels has been done by our clustering module similar to the example in Figure~\ref{fig:sample_patches}a. We propose an encoder-decoder model to learn annotations as shown in Figure ~\ref{fig:sample_patches}b. Clustered images were not pixel annotated, Figure ~\ref{fig:sample_patches}b is only for illustration. The architecture of the proposed model is shown in Figure \ref{fig:model}. The set of all training images is designated as $\mathbf{X}: \{\mathbf{x}_1, \mathbf{x}_2, ..., \mathbf{x}_N\}$. The encoder is designated: $q_\phi(\mathbf{x}_i)$ and the decoder is designated $p_\theta(\mathbf{z}_i)$. Additionally, $\mathbf{z}_i \in \mathbf{Z}$ is the corresponding latent vector to $\mathbf{x}_i$. The learning parameters in both encoder and decoder are $\phi$ and $\theta$ respectively, and $p_\theta(\mathbf{X})$ is the probability distribution of all seismic images over parameter $\theta$. Thus, the log-probability of the data is: $\log p_\theta(\mathbf{X})= \sum_i \log p_\theta(\mathbf{x}_i)$. Now, $p_\theta(\mathbf{x}_i)$ is intractable because the posterior $p_\theta(\mathbf{x}_i|\mathbf{z}_i)$ is intractable \cite[]{kingma2013auto}. For notational convenience, we will drop the subscript $i$ on $\mathbf{x}_i$ and $\mathbf{z}_i$, as all future references to these variables refers to a single data instance during iterative training or inference.

\hspace{8pt}We introduce two projection matrices: $\mathbf{P}_1$ and $\mathbf{P}_2$. Both matrices are fully connected layers of size 1024x1024. The matrices project $\mathbf{z}_i$ to orthogonal $\mathbf{z}_{1}$ and $\mathbf{z}_{2}$ subspaces respectively. In Figure~\ref{fig:sample_patches}b,  $\mathbf{z}_1$ and $\mathbf{z}_2$ are mapped to the blue and red annotations in each input image. Note that we do not assume there are no images with overlapping class representations in our clusters.  In one forward pass, operators $\mathbf{P}_1$ and $\mathbf{P}_2$ are applied to  $\mathbf{z}$ thus: 
\begin{equation}
    \mathbf{z}_1 = \mathbf{P}_1 \mathbf{z}, \; \; \; \mathbf{z}_2 = \mathbf{P}_2 \mathbf{z}.
\end{equation}

\noindent$\hat{\mathbf{x}}_1$ and $\hat{\mathbf{x}}_2$ are synthesized from the decoder, $p_\theta(\mathbf{z})$. The decoder tries to learn some properties of each seismic image and furnish such knowledge into reconstructing both images.
\begin{equation}
    \mathbf{\hat{x}}_1 \sim p_\theta(\mathbf{z}_1), \; \; \;
    \mathbf{\hat{x}}_2 \sim p_\theta(\mathbf{z}_2).
\end{equation}

\noindent We desire the reconstructed image $\mathbf{r}$ to be as similar to the input image as possible. Note that $\mathbf{r}$ is usually blurry in encoder-decoder networks. Hence, we use a discriminator, $D_1$,  to ensure $\mathbf{r}$ is sharp. A second discriminator, $D_2$, is applied to the encoder to ensure latent variables of $\mathbf{z}$ are spread out. In the original Bayes auto-encoder architecture, \cite{kingma2013auto} derive the evidence lower bound for reconstructing $\mathbf{x}$. Because we modify the behavior of the latent space, by projecting it to orthogonal latent spaces, we re-derive the evidence lower bound in the context of projecting the latent space. The decoder distribution, $p_\theta(\mathbf{X})$ is intractable because the posterior is intractable. Hence, we sample from an auxillary distribution, $q_\phi(\mathbf{z})$. We drop $\theta$ and $\phi$ for notation convenience knowing $p$ and $q$ are distributions over both parameters respectively. The derivation of the evidence lower bound (ELBO) on the log-likelihood of $\mathbf{X}$ is as follows:

\noindent $\mathbb{E}_{\mathbf{x} \sim p_{\mathbf{x}}} \; (\log \left(p_\theta(\mathbf{x})) \right)$ is log-likelihood of generating the data.
Next, we introduce $\mathbf{z}_1$ and $\mathbf{z}_2$ and sum over their marginals:

\begin{equation}
    \begin{aligned}
        \mathbb{E}_{\mathbf{x} \sim p_{\mathbf{x}}} \; (\log \left(p_\theta(\mathbf{x})) \right) \\ =  \mathbb{E}_{\mathbf{x} \sim p_{\mathbf{x}}} \left[ \log \left( \sum_{\mathbf{z}_1} \sum_{\mathbf{z}_2} p_\theta(\mathbf{x}, \mathbf{z}_1, \mathbf{z}_2)  \right) \right].
    \end{aligned}
    \label{marginals}
\end{equation}

\noindent We sample from $q(\mathbf{z|x})$ and find the expectation over its conditional distribution because we cannot sample from the posterior $p(\mathbf{x|z}))$: 
\begin{equation}
\begin{aligned}
    (\ref{marginals}) 
    \ge \mathbb{E}_{\mathbf{x} \sim p_{\mathbf{x}}} \left[ \mathbb{E}_{\mathbf{z}_1 \sim q(\mathbf{z}_1|\mathbf{x})} \; \log \left(\frac{p(\mathbf{z}_1)}{q(\mathbf{z}_1|\mathbf{x})} \right) \right.\\ +  \left. \mathbb{E}_{\mathbf{z}_2 \sim q(\mathbf{z}_2|\mathbf{x})} \; \log \left(\frac{p(\mathbf{z}_2)}{q(\mathbf{z}_2|\mathbf{x})} \right) \; \log \left(\frac{p(\mathbf{z}_2)}{q(\mathbf{z}_2|\mathbf{x})} \right) \right]\\ + 
    \mathbb{E}_{\mathbf{x} \sim p_{\mathbf{x}}} \left[ \mathbb{E}_{q(\mathbf{z}_1, \mathbf{z}_2)|\mathbf{x})} \; \log \left(p(\mathbf{x}|\mathbf{z}_1, \mathbf{z}_2)) \right) \right ]. \\
    \end{aligned}
    \label{auxillary}
\end{equation}
 
Simplifying expressions in (\ref{auxillary}) in $KL$ divergence terms gives:
\begin{equation}
    \begin{aligned}
        \mathbb{E}_{\mathbf{x} \sim p_{\mathbf{x}}} \left[ -KL(q(\mathbf{z}_1|\mathbf{x})||p(\mathbf{z}_1)) -KL(q(\mathbf{z}_2|\mathbf{x})||p(\mathbf{z}_2)) \right]  \\
+ \mathbb{E}_{q_{({\mathbf{z_1, z_2|x}})}}  \log(p(\mathbf{x|z_1,z_2})). \\
    \end{aligned}
    \label{KL_div}
\end{equation}

We can write the $KL$ terms in (\ref{KL_div}) in entropy terms:
\begin{equation}
    \begin{aligned}
        \mathbb{E}_{\mathbf{x} \sim p_{\mathbf{data}}} \left[ \mathbb{E}_{q_\phi(\mathbf{z_1, z_2|x})} \log \left( p_\theta(\mathbf{x|z_1, z_2}) \right) \right.\\  + 
        \left. H(q(\mathbf{z_1|x})) + H(q(\mathbf{z_2|x})) - H \left(p(\mathbf{z_1})\right)  -  H \left(p(\mathbf{z_2})
        \right)\right], 
    \end{aligned}
    \label{entropy}
\end{equation}

\noindent where $KL(\mathbf{a} || \mathbf{b})$ is the Kullback-Leibler divergence between distributions $\mathbf{a}$ and $\mathbf{b}$.
We could conclude from equation \ref{entropy} that: evidence $\ge$ reconstruction + entropy of projected latent codes - entropy of actual latent priors.
In equation \ref{entropy}, the entropy of the encoder is $H(q(\mathbf{z|x}))$ and the conditional entropy of the encoder due to the projected prior is $H(q(\mathbf{z_1|x}))$ and  $H(q( \mathbf{z_2|x}))$. The entropies of the priors in the orthogonal subspaces are: $H(p(\mathbf{z_1}))$ and $H(p(\mathbf{z_2}))$. If the conditional entropies of the the encoder after projecting the latent space matches the entropies of the projected priors: $H(q(\mathbf{z_1|x})) - H(p(\mathbf{z_1}))=0$ and  $H(q(\mathbf{z_2|x})) - H(p(\mathbf{z_2}))=0$, then the reconstructed image $\mathbf{r}$ will be a perfect reconstruction of $\mathbf{x}$. This would only happen if the projected priors are perfectly sampled from the actual priors of distributions representing the foreground and background of each $\mathbf{x}$. Two images are reconstructed by the decoder: $\hat{\mathbf{x}}_1$ and $\hat{\mathbf{x}}_2$. Here, we assume that the encoder learns not only to generate the latent codes, but understands that it needs to generate two latent codes that match the two virtual distributions in the decoder. Hence, it must generate one latent code to encapsulate two virtual distributions in the decoder. The decoder learns parameters $\theta$ such that each reconstructed seismic image has two factors inherent in them. It remains that the decoder must be constrained to understand both factors.

\subsection{Latent space factorization using projection matrices}
\label{proj_matrix}

\hspace{8pt}A projection matrix $\mathbf{P}$ is an $n \times n$ square matrix that gives a vector subspace representation $T$ that is a projection of $\mathbb{R}^n$ to $T$. For any real operator $\mathbf{P}$ to be a valid projection matrix, it must satisfy the following properties: 
\begin{enumerate}
    \item{$\mathbf{P} = \mathbf{P}^*$ ($\mathbf{P}^*$ is the adjoint of $\mathbf{P}$)}.
    \item{$\mathbf{P}^2 = \mathbf{P}$} (idempotent property).
\end{enumerate}
Figure \ref{fig:model} shows two projection matrices $\mathbf{P}_1$ and $\mathbf{P}_2$ designed to be mutually orthogonal.
However, we impose no constraint to make either of them orthogonal projection matrices in which case we would need $\mathbf{P} = \mathbf{P}^T$ for a real $\mathbf{P}$.
To impose a projection matrix behavior on both matrices, we create a fully connected layer in the model and initialize it using a uniform distribution. Now, both matrices are learnable and both projection and orthogonality constraints can be imposed on them by solving an optimization objective.
We formulate the following projection loss function:

\begin{equation}
    L_{proj} = \sum_{i,j, \; \; i \neq j}^2 \mathbf{P}_i^T \mathbf{P}_j + \sum_{i=1}^2 (\mathbf{P}^2_i - \mathbf{P}_i).
\label{lproj}
\end{equation}
Where $L_{proj}$ is the projection loss added to adversarial losses. We discuss the use of discriminators and their corresponding losses below.

\subsection{Adversarial Training}
\label{adversarial}
\hspace{8pt}\cite{goodfellow2014generative} introduce Generative Adversarial Networks (GANs). A GAN sets up an adversarial min-max game between a generator $G$ and a discriminator $D$. At training, the generator attempts to generate an image that lies in the dataset distribution space such that the discriminator is unable to distinguish if the image is real or generated. While the discriminator gets better at detecting real and generated (fake) images. This adversarial setup constrains the generator to generate sharp images almost at the quality of the input image. GANs are superior in the quality of generated images they produce compared to basic encoder-decoder models. Images reconstructed from encoder-decoder models are usually blurry. Blurry images are due to using the mean squared error (MSE) loss between the reconstructed and original image, implying the error between both images is Gaussian noise, which is not in the case in seismic images \cite[]{zhao2017towards}.  

For notation purpose, our decoder would pass as a generator $G$ in this section, while our encoder will be represented as $E$ in adversarial setting. The two discriminators specified in Figure \ref{fig:model} retain their notations as: $D_1$ and $D_2$. $D_1$ ensures a sharp reconstruction, $\mathbf{r}$, while $D_2$ helps in the spread of latent variables, $\mathbf{z}$. As seen in Figure \ref{fig:model}, $\mathbf{r} = \hat{\mathbf{x}}_1 + \hat{\mathbf{x}}_2$. We introduce an $L_1$ loss between $ \hat{\mathbf{x}}_1 \mbox{ and } \hat{\mathbf{x}}_2$ to enforce structural difference: 
\begin{equation}
    L_{diff} = -1 \times \sum_{n=1}^N |\hat{\mathbf{x}}_1 - \hat{\mathbf{x}}_2|.
    \label{ldiff}
\end{equation}

\noindent Where $L_{diff}$ is the $L_1$ penalty. Since discriminator $D_1$ takes $\mathbf{r}$ as input, the adversarial loss further ensure $\mathbf{r}$ is similar to $\mathbf{X}$. The adversarial loss for $D_1 \mbox{ versus } G$ becomes:
\begin{equation}
    \begin{aligned}
        \min_{G} \max_{D_1} & 
        E_{\mathbf{x} \sim p_{data}} \left[\log (D_1(\mathbf{x})) \right] \\ +   E_{\mathbf{z} \sim q(\mathbf{z})} \left[ \log(1-D_1(G(\mathbf{z}))) \right].
  \end{aligned}
    \label{ldav1}
\end{equation}
\noindent Similarly, we define an adversarial loss on the latent space $\mathbf{z}$. 

Training a GAN is very challenging. One experimental problem with training GANs is the problem of mode--collapse. Mode--collapse occurs when $G$ reconstructs the same output image for varying latent variable $\mathbf{z}$. In which case, $\mathbf{r}$ converges to a local minimum of one or few images without generalizing properly on the whole data distribution. To solve this problem, we apply discriminator $D_2$ on $\mathbf{z}$. $D_2$ helps to impose a uniform distribution on $\mathbf{z}$ and to flatten out the mixed Gaussian space of $\mathbf{z}$.  
\begin{equation}
    \begin{aligned}
        \min_{G} \max_{D_2} & 
        E_{\mathbf{z} \sim q(\mathbf{z})} \left[\log(D_2(\mathbf{z})) \right] + E_{\mathbf{u} \sim U_{[0,1]}} \left[\log(1-D_2(\mathbf{u})) \right]. \\
    \end{aligned}
    \label{ldav2}
\end{equation}
Equation \ref{ldav2} defines the loss on $D_2$, to flatten the mode of the Gaussian distribution in the latent space. Thus, the mode-collapse problem in the latent space is diminished. Equations \ref{ldav1} and \ref{ldav2} are defined as $L_{D_1} \mbox{ and } L_{D_2}$ respectively.

\subsection{Reconstruction Loss}
\hspace{8pt}In equation \ref{ldiff}, we impose a structural difference loss between $\hat{\mathbf{x}}_1$ and $\hat{\mathbf{x}}_2$. We need a reconstruction loss to ensure both images do not become a trivial solution while minimizing $L_{diff}$ in equation \ref{ldiff}. A trivial solution to $L_{diff}$ could be $\hat{\mathbf{x}}_1=\mathbf{1}$, a matrix of all 1s, and  $\hat{\mathbf{x}}_2 =\mathbf{0}$, a matrix of all 0s. To avoid this, we define a reconstruction loss on $\mathbf{r}$ as follows:

\begin{equation}
    L_{reconst.} = \frac{1}{N} \sum_{n=1}^N |{\mathbf{x}} - \mathbf{r}|^2.
\end{equation}
\noindent But $\mathbf{r} = \hat{\mathbf{x}}_1 + \hat{\mathbf{x}}_2$. Hence,

\begin{equation}
    \label{reconstruction}
    L_{reconst.} = \frac{1}{N} \sum_{n=1}^N |{\mathbf{x}}_n - (\hat{\mathbf{x}}_{1_n} + \hat{\mathbf{x}}_{2_n})|^2.
\end{equation}

\noindent Equation \ref{reconstruction} ensures both $\hat{\mathbf{x}}_1$ and $\hat{\mathbf{x}}_2$ are valid seismic images and ensures a cycle consistency with the input image. 

Lastly, the  composite  model is trained  with  losses (\ref{lproj}-\ref{reconstruction}) over 300 epochs. After  each  forward pass,  we  back-propagate $L_{reconst}$ to update $E$, $G$, $\mathbf{P}_1$, $\mathbf{P}_2$. $L_{D_1}$ is back-propagated to update models $D_1$ and $G$. While $L_{D_2}$  is back-propagated to update models $D_2$ and $E$. Finally, $L_{proj}$ and $L_{diff}$ losses are used to update $\mathbf{P}_1$ and  $\mathbf{P}_2$ simultaneously.

\section{Results}
\label{results}

\begin{figure*}[htb!]
\centering
\begin{tabular}{cc}
  \includegraphics[scale=0.54]{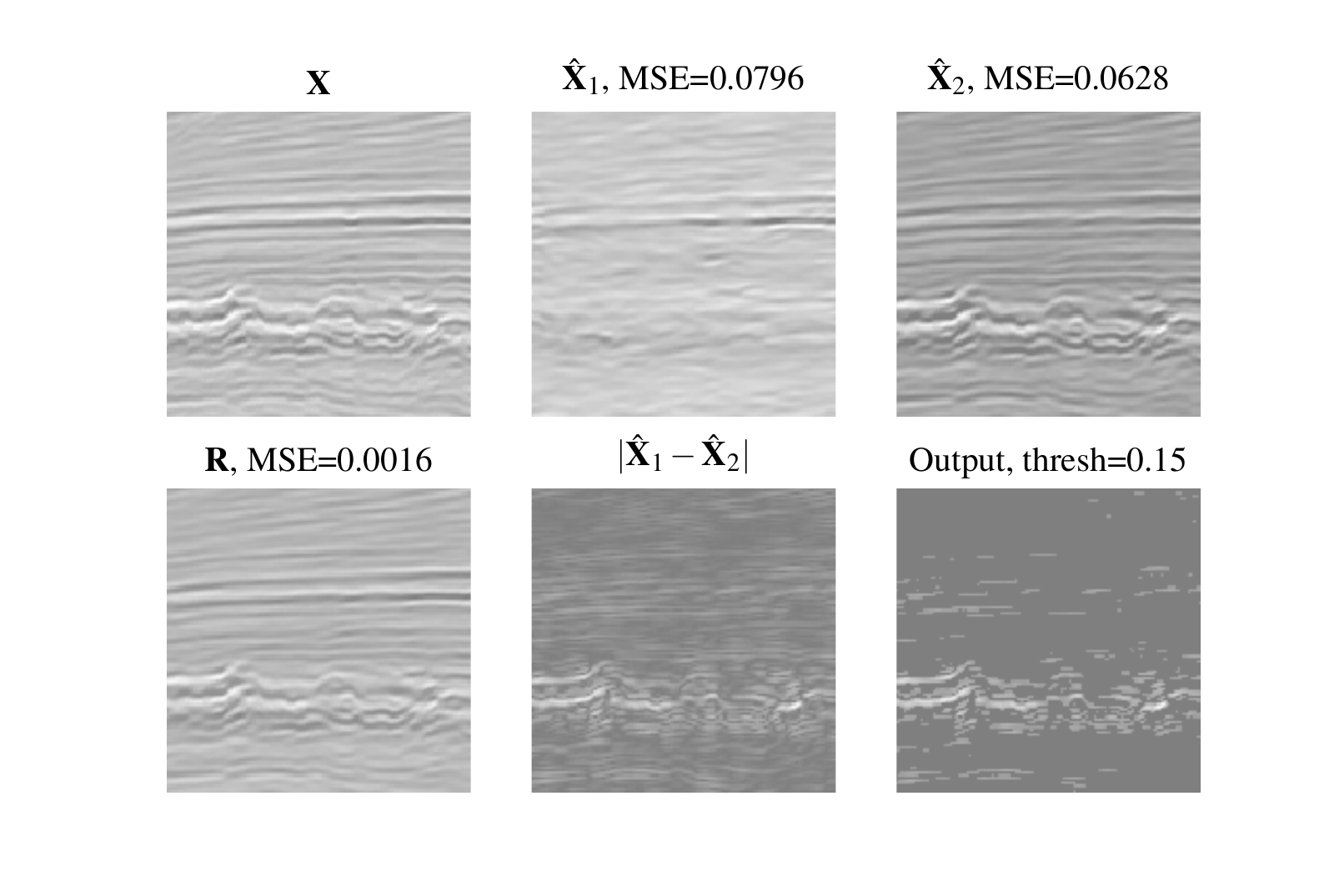} &   \includegraphics[scale=0.54]{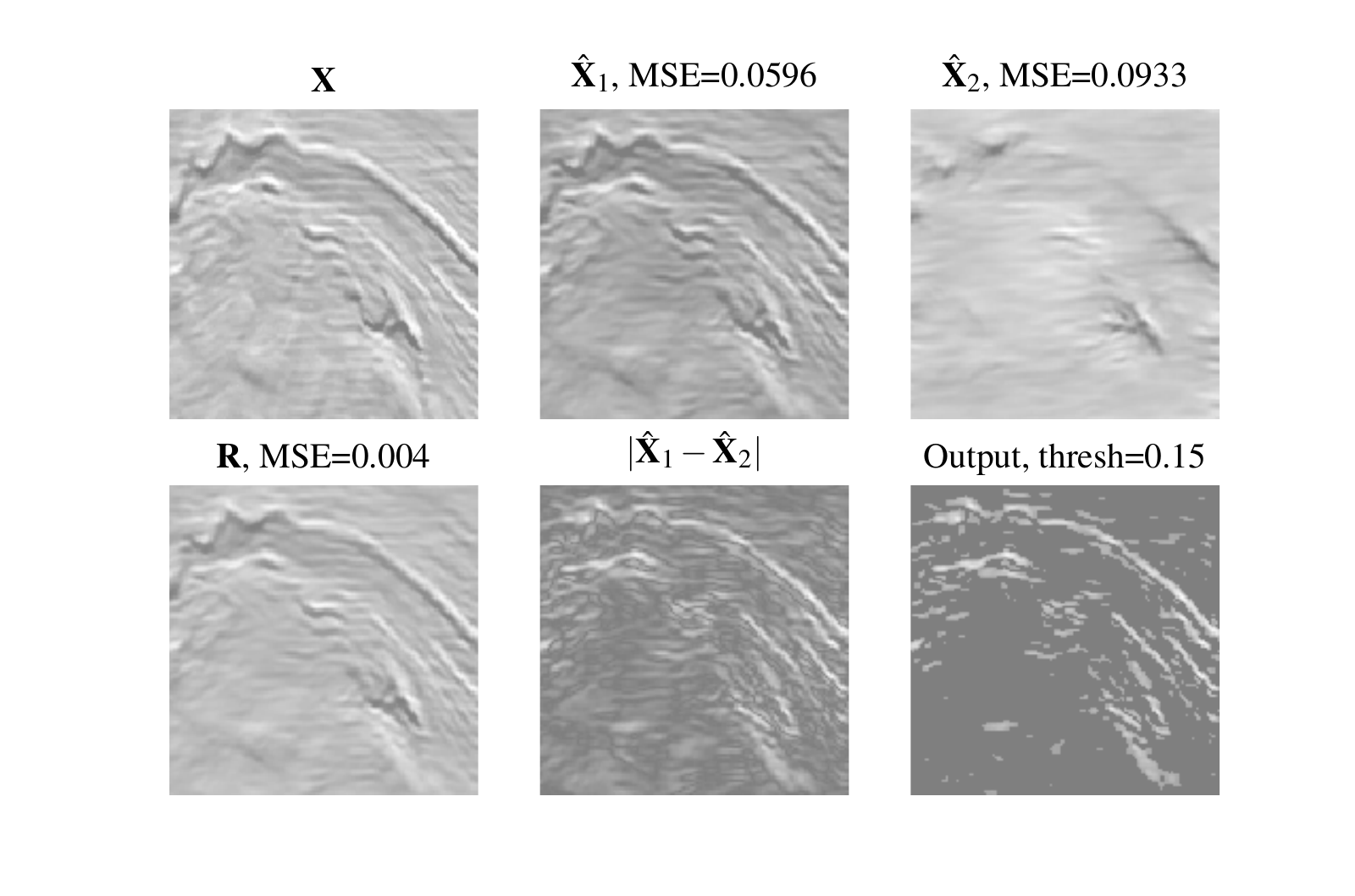} \\
(a) Chaotic Region Delineation & (b) Salt Region Delineation \\
 \includegraphics[scale=0.54]{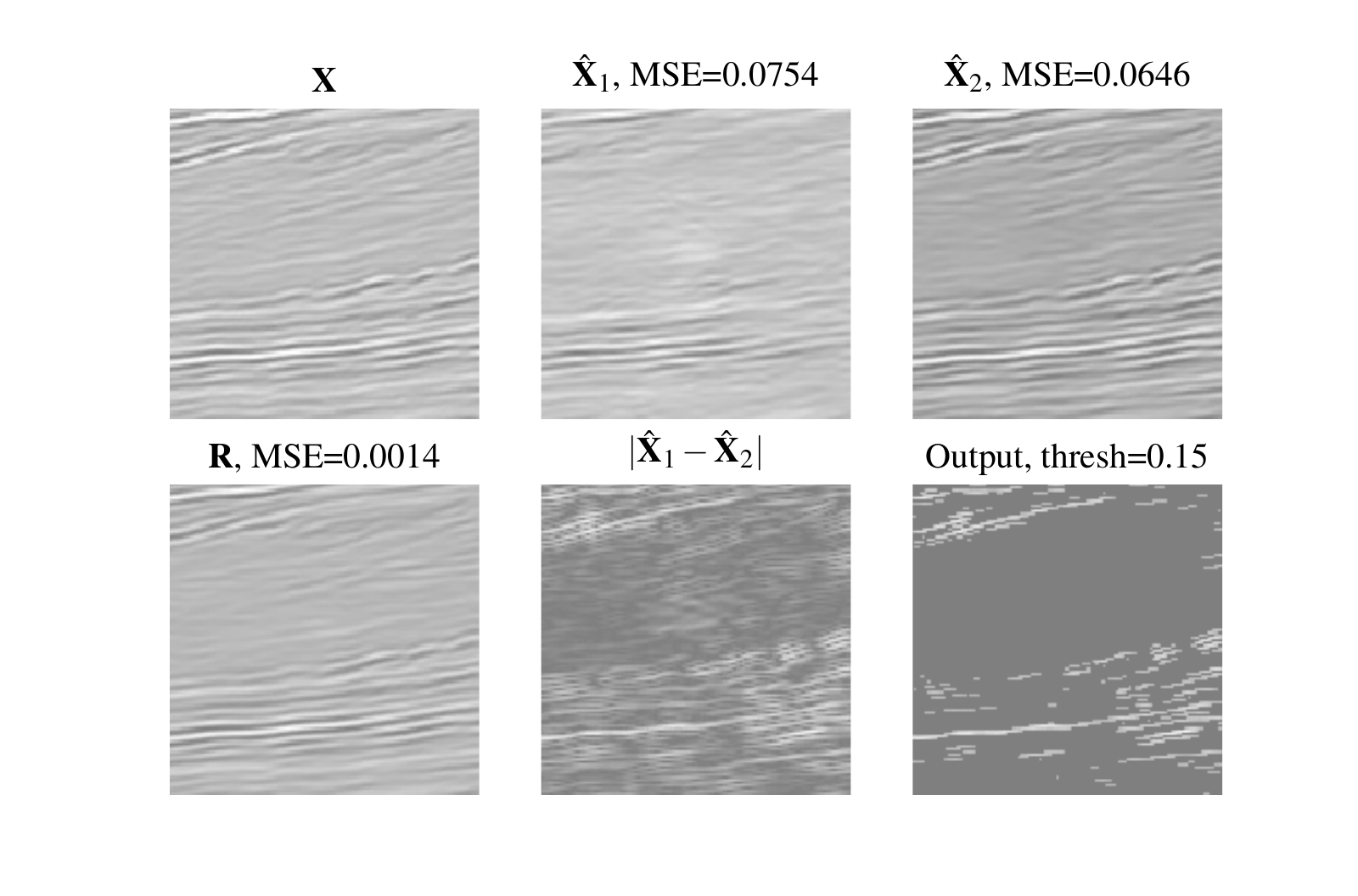} &   \includegraphics[scale=0.54]{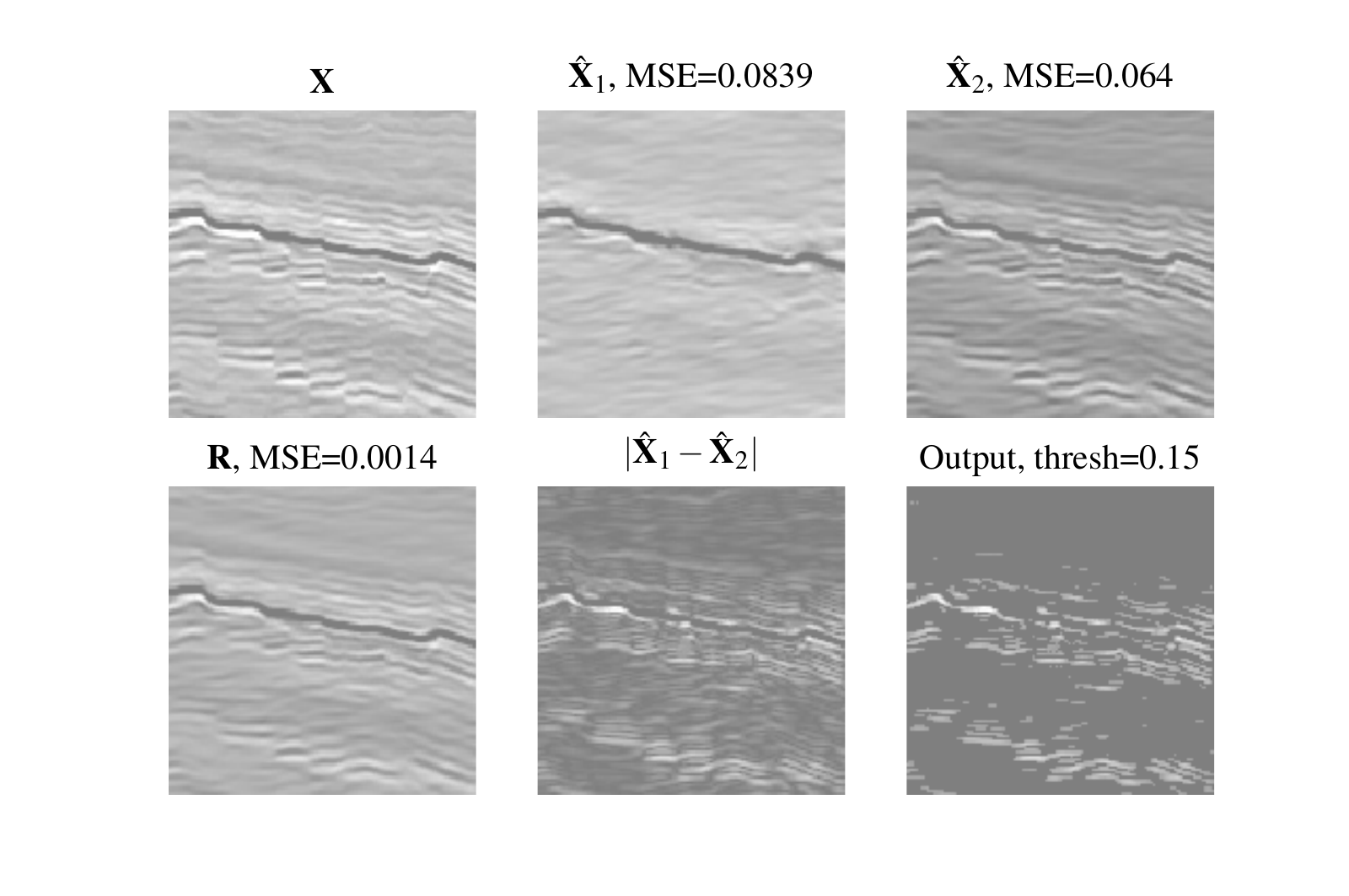} \\
(c) Horizon Lines Delineation & (d) Fault Region Delineation \\
\end{tabular}
\caption{Proposed output images from our latent space factorization model trained on 120x120 patches. In each sub-figure, there are six images. $\mathbf{x}$ is the input image. $\hat{\mathbf{x}}_1$ is the image synthesized from projection matrix $\mathbf{P}_1$ and likewise $\hat{\mathbf{x}}_2$ is synthesized from $\mathbf{P}_2$.  $\mathbf{r}$ is the reconstructed image. $|\hat{\mathbf{x}}_1-\hat{\mathbf{x}}_2|$ is the $L_1$-norm sparse output obtained from minimizing $L_{diff}$ in equation \ref{ldiff}. (a) shows the delineation of chaotic facies in the image. For each output image, we threshold out pixel values below 0.15. (b) shows a patch from a salt region. Notice the sparse pixel labels identifies interesting structures in the image. (c) shows parallel reflectors are delineated differently from the regions of smoothed amplitude reflections. (d) shows step-wise amplitude reflectors in the faulted region. Our method does not delineate faults along the faulting planes because it is unsupervised, but this regional delineation is sufficient to extend our patch model to sectional volume fault region delineation. Lastly, $\hat{\mathbf{x}}_1$ in (a), (c), (d) and $\hat{\mathbf{x}}_2$ in (b) were synthetically generated by our algorithm to satisfy our factorization objective because they are dissimilar to images in our training set.}
\label{patch-delineation}
\end{figure*} 

\hspace{8pt}We show qualitative results on the trained images and generalize our predictions on the output images to make predictions on full vertical seismic sections. In addition, we show how our method can be applied to attribute extraction. In our attribute extraction process, we analyze two complementary attributes and compare them to six existing attributes in literature.

\hspace{8pt}First, each output image from the entire workflow has both an assigned class and a pixel label corresponding to a region of interest. In Figure~\ref{patch-delineation}, the input image $\mathbf{x}$ obtained from our clustering module is passed to the latent space model to obtain $\hat{\mathbf{x}}_1$ and $\hat{\mathbf{x}}_2$. In each sub-figure of Figure~\ref{patch-delineation}, $\hat{\mathbf{x}}_1$ and $\hat{\mathbf{x}}_2$ are synthesized by the model to be structurally complementary to each other such that their combination reconstructs $\mathbf{x}$.  And the $L_1$ norm of their difference reveals a geologically meaningful region. In each sub-figure, we show sparse inference regions using confidence values. Each pixel value is a probability value in $[0, 1]$, where $0$ is the probability a pixel is not part of a geological class and $1$ is otherwise. For improved performance, we threshold-out probability values below a threshold to obtain a cleaner binary image at the bottom right. An evidence to show that both $\hat{\mathbf{x}}_1$ and $\hat{\mathbf{x}}_2$ are factorized from $\mathbf{x}$, as we hypothesized, is to observe their mean-squared error (MSE) against input image $\mathbf{x}$ .
In all four sub-figures, both synthesized images ($\hat{\mathbf{x}}_1$, $\hat{\mathbf{x}}_2$), have higher MSEs than the reconstruction $\mathbf{r}$. This implies that both images are not as similar to the input image as the reconstructed one, because they are factorized components of the input image. Further evidence is  observed in $\hat{\mathbf{x}}_1$ in Figures \ref{patch-delineation}(a), \ref{patch-delineation}(c), \ref{patch-delineation}(d) and $\hat{\mathbf{x}}_2$ in Figure \ref{patch-delineation}(b). 
In all these instances, the model generates an image that is unlike any image in our training set. The orthogonal counterpart of these images are similar to our input images with more contrast. Hence, $\hat{\mathbf{x}}_1$ \mbox{ and } $\hat{\mathbf{x}}_2$ are equivalent representation of the orthogonal subspaces to which their latent subsidiaries were projected. 

\begin{figure*}[htb!]
    \centering
    \includegraphics[width=\textwidth]{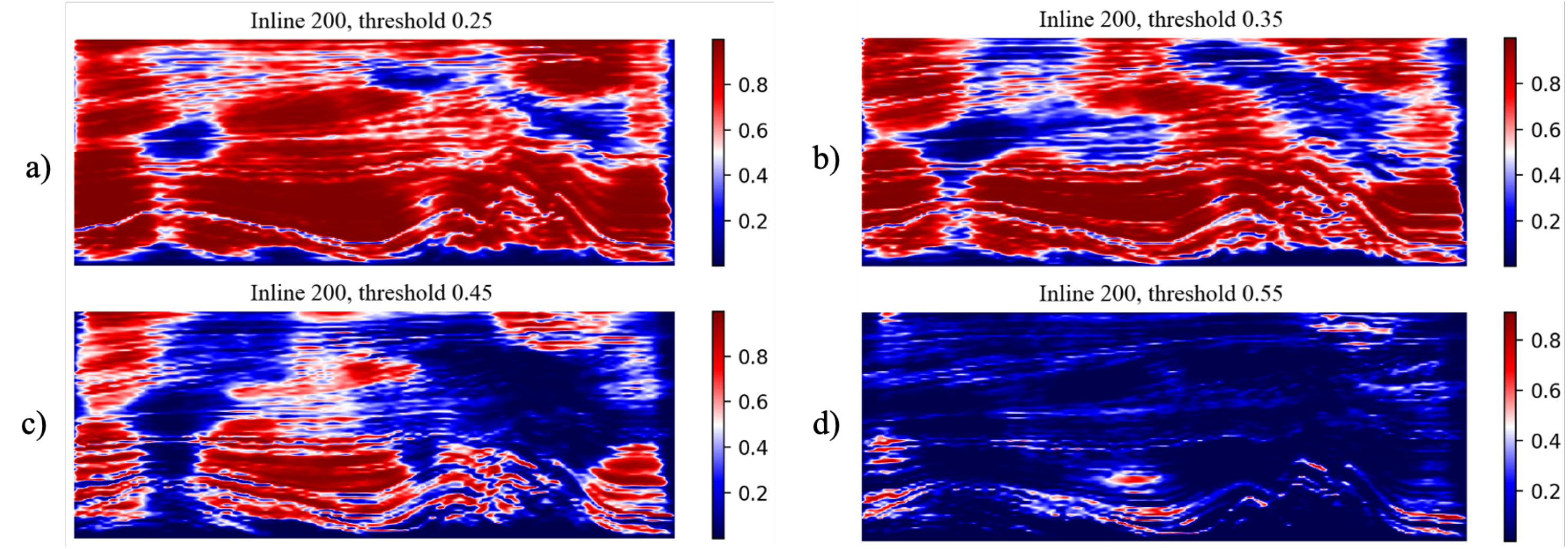}
    \caption{This figure shows effect of four cut-off thresholds on predicted sections. a) with threshold set to 0.25, the prediction is quite noisy. b) when the threshold is set to 0.35, the prediction is less noisy. c) shows further progressive improvement in prediction at threshold=0.45. d) shows clear structural prediction when the threshold when set to 0.55.}
    \label{fig:thresholds}
\end{figure*}

The image label and pixel annotation obtained in Figure~\ref{patch-delineation} can be generalized to label sections of the F3 block to assist with seismic interpretation. We trained a segmentation model - DeepLab V3+ \cite[]{chen2017deeplab} on all output images from our proposed pipeline and applied a soft-max classification layer to make a prediction on each reflection amplitude to any of the pre-defined four classes. Furthermore, we ran another experiment to determine whether each amplitude in the seismic section belongs to a structure class. The later experiment fits into an attribute extraction framework to be discussed later. In all predicted pixel labels, we threshold out small probability values to improve our confidence on the pixel labels. Figure~\ref{fig:thresholds} show four levels of threshold values:  $0.25, 0.35, 0.45$ and $0.55$. 

As shown in Figure~\ref{fig:thresholds}, images with a cut-off of $0.55$ performed best on structural classification of the seismic amplitudes when trained on DeepLab V3+ model. Hence, we report results based on a threshold of $0.55$. Note that the colorbars in Figure~\ref{fig:thresholds} do not reflect the thresholds. The thresholds are applied to the output images of the latent space model which are then used to train the segmentation network.
                                    
\begin{figure*}[htb!]
    \includegraphics[width=\textwidth]{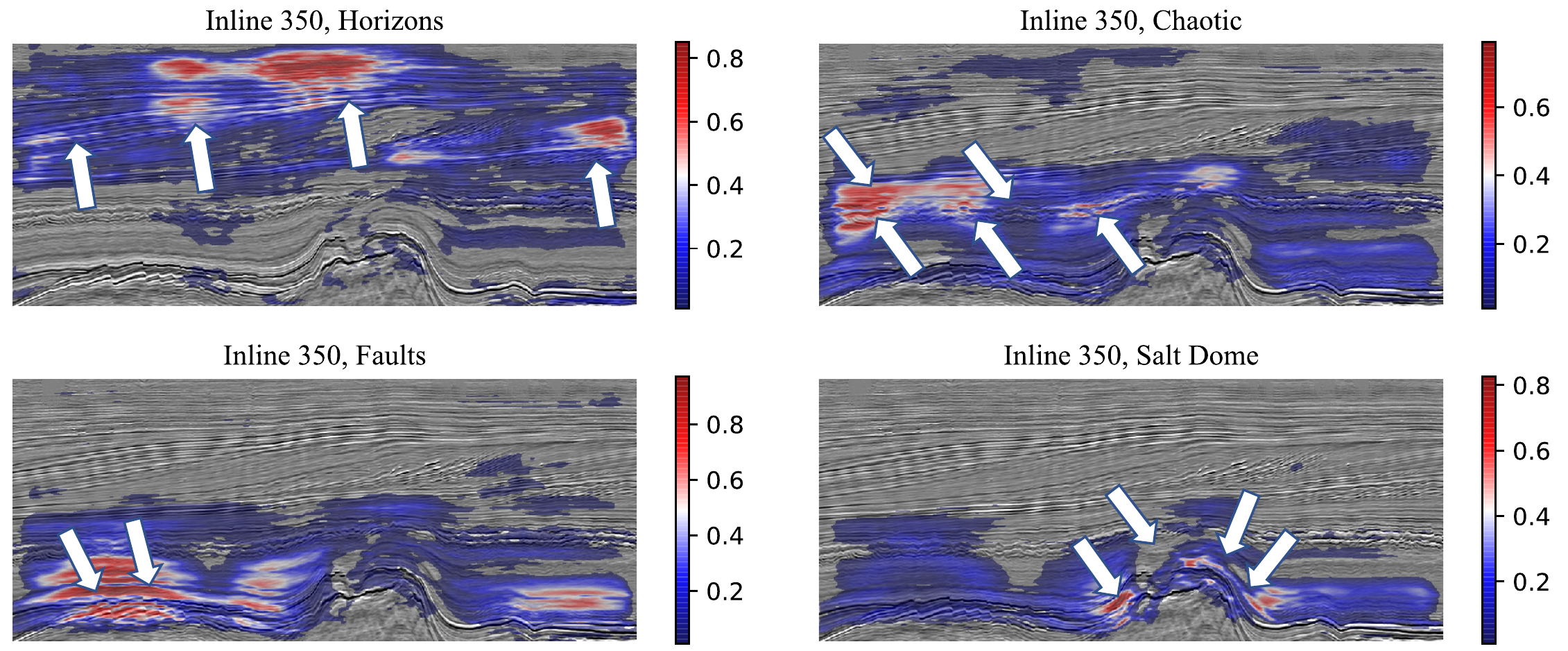}
    \caption{Factorized 2D sections of the F3 Block showing a) horizons, b) chaotic, c) faults and d) salt dome structures in inline 350 of the F3 block. The white arrows illustrate regions delineated correctly by our framework}
    \label{fig:2d}
\end{figure*}

\hspace{8pt}Figure~\ref{fig:2d}, shows the corresponding results for a test section at inline 350 of the F3 block. Figure~\ref{fig:2d}a identifies parallel reflectors with dipping amplitude reflections. In Figure~\ref{fig:2d}b, an approximate blurb is shown on the chaotic sedimentary strip of the section. The chaotic region to the right is not as well delineated as the region on the left. A possible explanation for this observation is that most of the images clustered into the chaotic class were taken from regions on the left of the F3 block. If the chaotic facies on the right differs from the one on the left, then this partial delineation is a derived anomaly that may be corrected by including images from the right chaotic region into our training set. However, since our method of model training is self-supervised, we report our result as is. 
Figure~\ref{fig:2d}c identifies the region containing fault structures on the left of the salt dome in the highlighted section. The highlighted region on the right of the salt dome is a false positive. However, the left identified region is delineated correctly. Note that the model delineates regions with faults without tracking the fault lines. Delineating regions of faults helps interpreters locate fault regions from which manual tracking of the faults may be done. Although salt domes are challenging to delineate, we show that our method identifies the salt dome boundary with significant accuracy in Figure~\ref{fig:2d}d. Other non-salt regions to the left and right of the salt are lightly highlighted. They are also false positives but we can dismiss them due to low probability values ($\sim 0.25$) in those regions. 

\begin{figure*}[htb!]
    \centering
    \includegraphics[width=\textwidth]{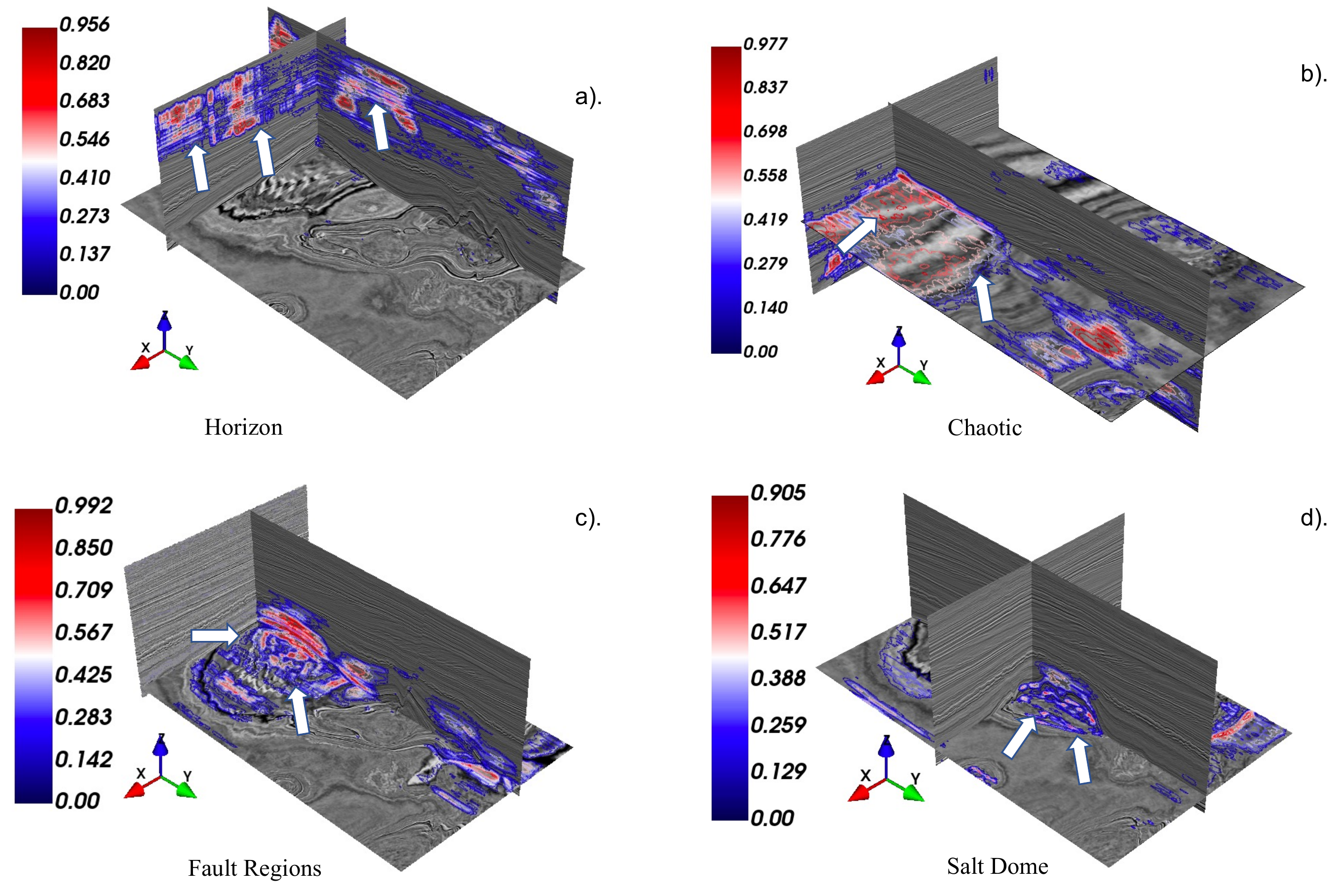}
    \caption{3D Delineation of structures in the F3 Block. The $z$-axis is the time axis, the x-axis is the inline direction and the y-axis points in the crossline direction. a) shows parallel reflector lines delineated, b) shows a vertical section across the chaotic strip delineated, c) shows regions with fault structures delineated, d) shows the salt-dome region region delineated by our framework.}
    \label{fig:3d}
\end{figure*}

We extended our section-based structural delineation in Figure~\ref{fig:2d} to 3D volume factorization of the F3 block. Figure~\ref{fig:3d} shows four facies in four separate classification instances in the F3 block. The delineated facies are marked using a white arrow. Evidence that factorization of the volume occurs can be seen by the relative absence of other structures apart from the classified one. For instance in Figure~\ref{fig:3d}a, only parallel reflectors from the horizon class are shown while the salt dome, fault region, and chaotic facies are factorized out. In Figure~\ref{fig:3d}b, the chaotic facies is delineated and the parallel reflectors shown in Figure~\ref{fig:3d}b are absent. In fault regions shown in Figure~\ref{fig:3d}c, the delineated region the right is a false positive corresponding to the false positive identified region on the right of the salt dome in Figure~\ref{fig:2d}. However, the left white arrows identify the fault region. Note that in Figure~\ref{fig:3d}d, the salt dome is delineated, but the horizons and chaotic regions are factorized out. The 3D delineation applies mostly to the boundary of the salt dome structure and corresponds to the regions delineated in Figure~\ref{fig:2d}d.

\subsection{Attribute Extraction}
\label{attributes}
\hspace{8pt}We further demonstrate that our deep learning model can factorize the F3 block into foreground and background attributes, each acting as features to guide the seismic interpreter to make informed decisions. The DeepLab segmentation model is modified to classify each amplitude in a section into two classes: foreground and background. Hence, every amplitude is mapped to a binary class. For each section, we compare the two predicted classes to six attributes from the literature for qualitative assessment. Gradient of texture (GoT) is a recent seismic attribute based on the gradients of seismic amplitudes in a moving window. In the 3D version, the gradients in the x, y and z directions were measured and combined into one value per pixel. The 2D version calculates gradients only in the x, y coordinates. The GoT attribute has been applied to delineate the boundaries of salt domes in 2D sections and 3D seismic volumes \cite[]{shafiq2015detection}. We compared our proposed attributes to 2D and 3D GoT.

\begin{figure*}[htb!]
    \centering
    \includegraphics[width=\textwidth]{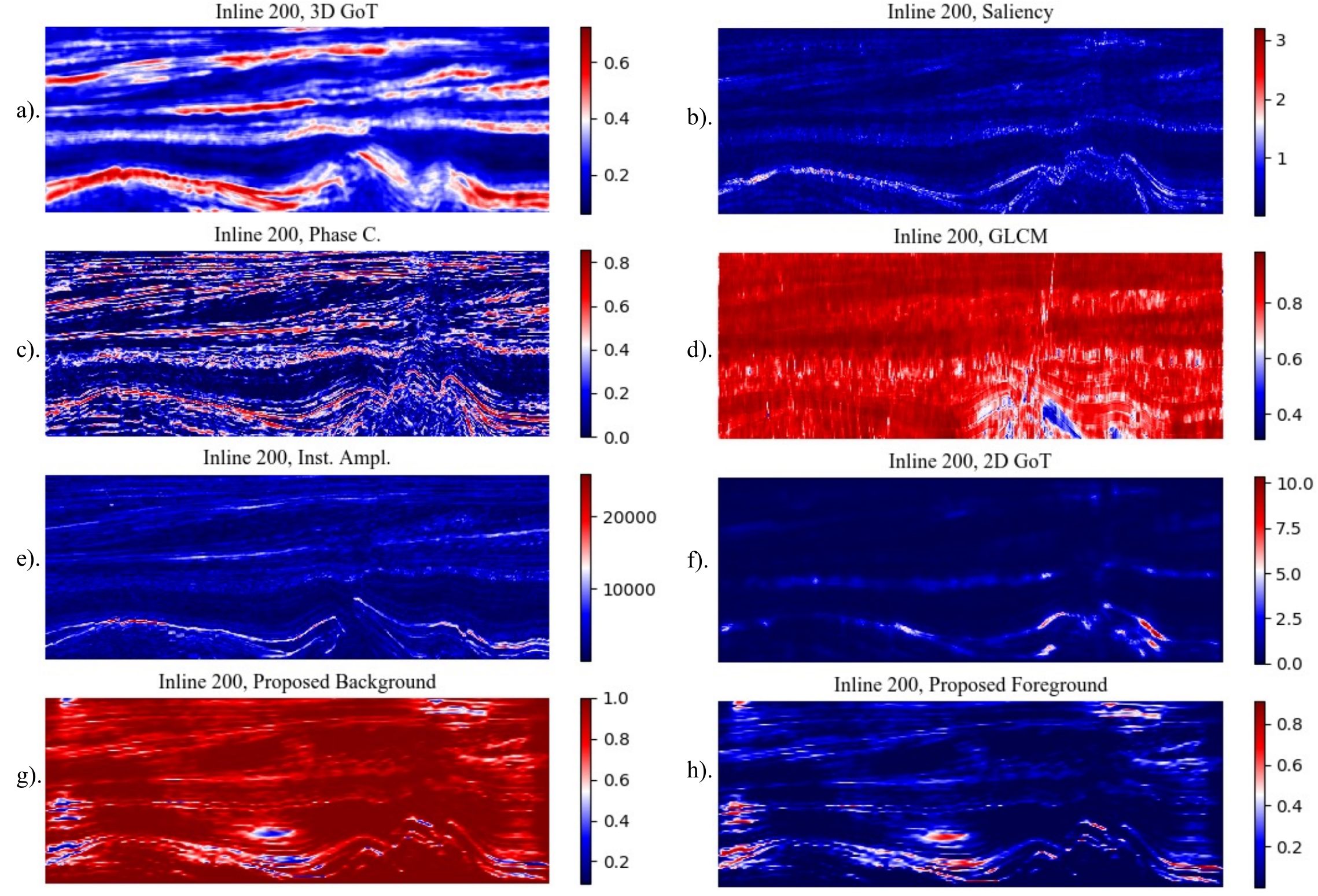}
    \caption{This figure shows eight sub-figures of various attributes extracted from the same F3 block volume. All attributes are shown for inline section 200. a) is 3D GoT attribute shown for inline 200. b) is the human visual system based saliency map for the same section, c) shows phase congruency. Note that phase congruency is an improvement to 3D-GoT. d) is Texture-based GLCM. GLCM performs poorly for structural delineation in this application. e) is the Instantaneous amplitude attribute. This attribute identifies the boundary of the salt dome and highlights a few horizon lines. For completion, we include 2D GoT in f). 2D GoT does not reveal many features aside from the edges of the salt dome. g) and h) are our proposed background and foreground-based attributes. Notice that in h), most features we seek are represented. The salt dome boundary, the chaotic strip, and the horizon lines. The fault region is lightly delineated. g) is complementary to h). It shows the background regions that do not belong to features identified as structures.}
    \label{fig:attributes}
\end{figure*}

Phase congruency \cite[]{shafiq2017salt} is another recent seismic attribute that improved on GoT. Phase congruency is a dimensionless quantity and it is unaffected by changes in image illumination and contrast, unlike GoT. 
  
\hspace{8pt}In addition, we compared our method against a saliency-based seismic attribute \cite[]{shafiq2018role}, which view the features in the seismic volume in a perception analogous to human vision. The component features of saliency attributes consist of fast Fourier Transform (FFT) coefficients. Similar to 3D GoT, saliency-based attributes can be computed in 2D and 3D variants. Lastly, we compare against gray-level co-occurrence matrices (GLCM) texture attributes \cite[]{chopra2006applications} and instantaneous amplitude  \cite[]{white1991properties} attributes. 
  
Figures~\ref{fig:attributes}g and \ref{fig:attributes}h are our proposed attributes, while Figures \ref{fig:attributes}a - \ref{fig:attributes}f are attributes we compared against.  Figure~\ref{fig:attributes}g is the proposed background attribute and Figure~\ref{fig:attributes}h is the proposed foreground attribute. GoT and Phase congruency, in Figure~\ref{fig:attributes}a and \ref{fig:attributes}c, show better delineation of parallel reflectors than our foreground attribute shown in Figure~\ref{fig:attributes}h. However, our method shows a less noisy delineation of the parallel reflectors and highlights the most important ones. Saliency, instantaneous amplitude and 2D-GoT in Figure~\ref{fig:attributes}b, \ref{fig:attributes}e, and \ref{fig:attributes}f delineates mostly the salt boundary and the fault regions - which implies our framework performs better in delineating parallel reflectors. GLCM at Figure~\ref{fig:attributes}d performed least in delineating structural attributes among all eight attributes compared.

\subsection{Comparison with a Machine Learning Framework by \cite{alaudah2018structure}}
\label{compare_aludah}
\begin{figure*}[htb!]
    \centering
    \includegraphics[width=\textwidth]{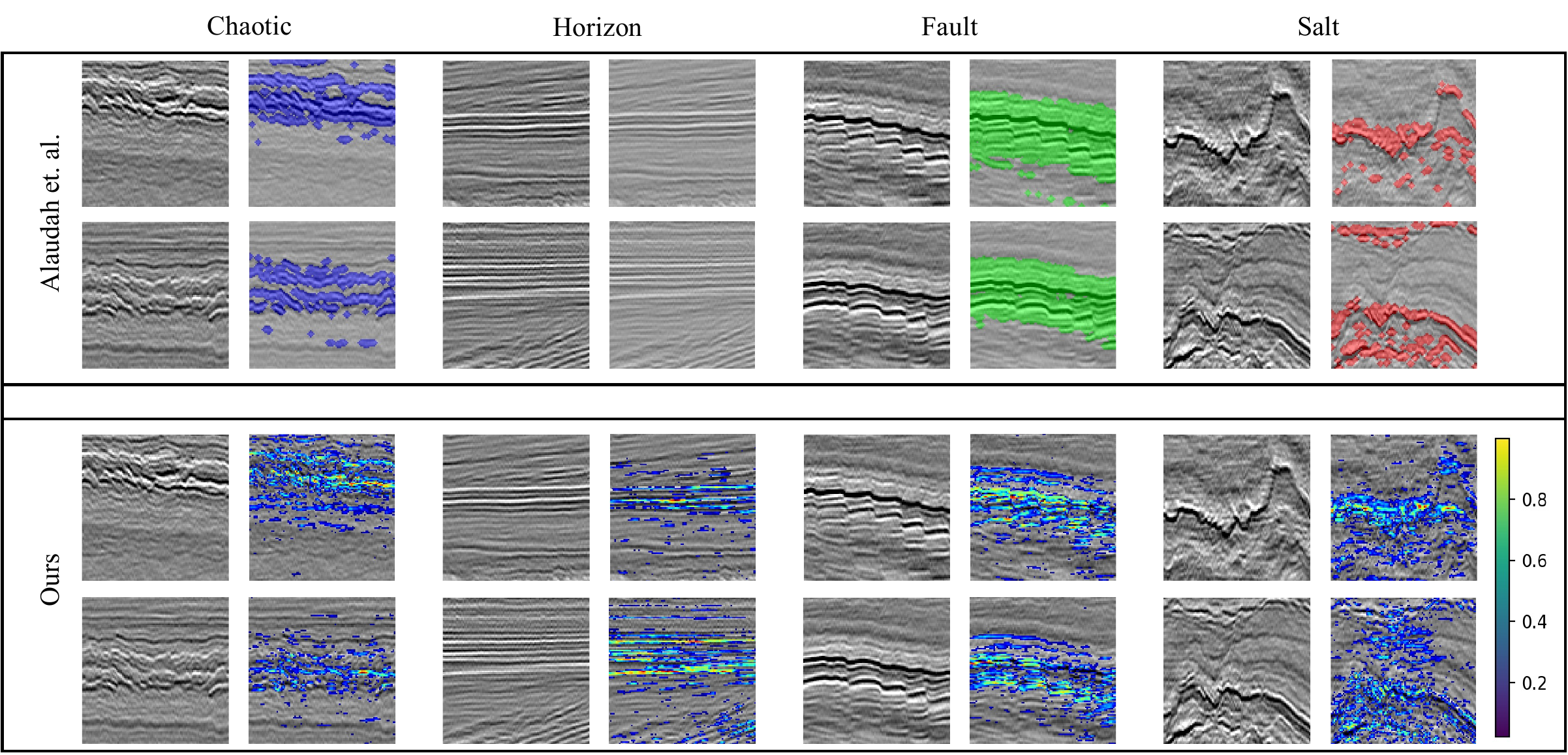}
    \caption{We make a class-by-class comparison of our unsupervised model to \cite{alaudah2018structure}'s weakly supervised model.}
    \label{fig:alaudahvsus}
\end{figure*}

\hspace{8pt}Our deep learning framework could also be applied to solve a label-mapping problem similar to the problem solved by \cite{alaudah2018structure}. The author applied a non-negative matrix factorization (NNMF) algorithm to predict pixel labels from image labels. The NNMF factorizes an input matrix $X$ into features and label assignments: $$X = WH,$$ where $X$ is a matrix with images as column entries. $W$ is a features vector and $H$ gives the assignment of features in $W$ into their respective classes. The features learned during factorization were mapped to corresponding images to delineate geological structures. In Figure~\ref{fig:alaudahvsus}, we attempt to label pixels by mapping image labels learned from our clustering framework to pixel predictions made by our deep learning model and we compare the result with \cite{alaudah2018structure}'s framework. 

The author used four classes: chaotic, salt dome, faults, and others, where the latter include images that do not belong to the previous three classes. Two major differences between our framework and the author's framework are as follows: in the latter, an interpreter assigns image labels to a few examples, uses an image retrieval technique to obtain labels for the remaining images which is then applied to the weakly-supervised framework. Secondly, \cite{alaudah2018structure}'s method is not based on a deep learning framework.  In contrast, we do not use image labels and we train a deep learning frame for pixel and image labels. Figure~\ref{fig:alaudahvsus} shows four classes of images. Note that in labeling pixels in all the classes, the accuracy of our framework is highest along edges while \cite{alaudah2018structure}'s labeling is more region oriented. For instance, our framework assigns the highest confidence values of delineated geologic features to edges of peaks-to-troughs in all four classes. This is because we used an $L_1$-norm sparsity loss in generating the probability maps for predicting structures. In the chaotic features delineated, the performance of both frameworks is close. Because \cite{alaudah2018structure}'s method was not trained on the horizon's class specifically, the delineation of structures in this class of images was left out by the algorithm. However, in the same class, we are able to delineate lines of parallel reflectors.

\cite{alaudah2018structure}'s labeling captured the fault region more elegantly than ours, but our method highlights the salt regions elaborately than \cite{alaudah2018structure}'s method which mostly labeled the edges or boundary of the salt structure. Salt imaging is challenging due to the rugose structure of salt domes, leading to complex amplitude patterns in salt facies. These complex patterns can conflict with other structural patterns in the feature factorization of \cite{alaudah2018structure}'s framework leading to relatively poorer performance. Our method focuses attempts to separate interesting features from the surrounding background, hence it captures complex patterns better than the previous framework. 

\section{Conclusion}
\hspace{8pt}We proposed two frameworks based on hierarchical clustering and a deep adversarial model. We showed that our hierarchical clustering model can be used for unsupervised clustering of seismic images. We also showed that our self-supervised deep learning framework can be used for self-supervised segmentation of geological structures using orthogonal latent space projection. One limitation of our self-supervised model is that the delineated structures are partial, which leaves room for improvement. Furthermore, the hierarchical clustering part and the self-supervised part are separate modules. In future work, we could create a comprehensive framework that would not need a pre-clustering module. The adversarial training method could also be improved for more stable training. Lastly, the latent space factorization methodology could be extended to multiple geological component delineations in each image compared to the current foreground and background-based methodology.

\section{Appendix - A Proof of Equation \ref{entropy}}

\hspace{8pt}Here, we provide detailed proof of equation \ref{entropy}.
Let $\mathbf{X}$ be our input data of images and $\mathbf{X} = [\mathbf{x}_1, \mathbf{x}_2, ..., \mathbf{x}_N ]$. We can assume independence on each $\mathbf{x}_i$. Let the distribution of $\mathbf{X}$ be $p_{data}$. Assume there exists a decoder/generator from which we can generate $\mathbf{x} \in \mathbf{X}$. Let the distribution of this generator over some learned parameter $\theta$, be $p_{\theta}(\mathbf{x})$. The log-likelihood of generating $\mathbf{X} = \mathbb{E}_{\mathbf{x} \sim p_{data}}[\log(p_\theta(\mathbf{x}))]$. We introduce an auxiliary distribution over another learning parameter $\phi$, $q_{\phi}(\mathbf{x})$ to approximate $p_{\theta}(\mathbf{x})$. We introduce $q_\phi(\mathbf{x})$ because to reconstruct each $\mathbf{x}$ from the  likelihood function, we must know the underlying distribution. We can re-write the log-likelihood as:

\begin{equation}
    \begin{aligned}
        \mathbb{E}_{x \sim p_{data}} \left[ \log 
        \left( p_\theta( \mathbf{x}) \right) \right] \\ = \mathbb{E}_{x \sim p_{data}} \left[ \log 
        \left(\sum_\mathbf{z}  p_\theta( \mathbf{x, z}) \right) \right] \\
    \end{aligned}
\end{equation}

\begin{equation}
    \begin{aligned}
        & = \mathbb{E}_{x \sim p_{data}} \left[\log \left( \sum_\mathbf{z} p_\theta(\mathbf{x|z}) . p_\theta(\mathbf{z}) \right) \right],
    \end{aligned}
\label{eqn:appdx}
\end{equation}

\noindent but the posterior $p_\theta(\mathbf{x|z})$ is intractable, hence we use an auxiliary distribution, $q_{\phi}(\mathbf{x})$, to estimate $p_\theta(\mathbf{x|z})$. Note that we introduce a prior $\mathbf{z}$ into equation \ref{eqn:appdx}. From an encoder-decoder framework, the prior $\mathbf{z}$ is the latent space variable passed into the decoder/generator. We know the distribution of $\mathbf{z}$ is Gaussian with zero mean, unit variance due to batch normalization at the end of the encoder. Hence, we re-write equation \ref{eqn:appdx} as:
\begin{equation}
    \begin{aligned}
    \noindent\mathbb{E}_{\mathbf{x} \sim p_{data}} \left[\log \sum_\mathbf{z} (p_\theta( \mathbf{x, z})) \right] 
    \end{aligned}
\end{equation}
we can condition the likelihood on $\mathbf{z_1, z_2}$ without loss of generality:
\begin{equation}
    \begin{aligned}
    & = \mathbb{E}_{\mathbf{x} \sim p_{data}}\left[ \log \left(\sum_\mathbf{z_1} \sum_\mathbf{z_2} (p_\theta(\mathbf{x|z_1, z_2})) . p_\theta(\mathbf{z_1, z_2}) \right) \right].
    \end{aligned}
    \label{lgen}
\end{equation}
Next, we introduce $q_\phi(\mathbf{z_1}|\mathbf{x})$, $q_\phi(\mathbf{z_2}|\mathbf{x})$ into (\ref{lgen}) follows:
\begin{equation}
    \begin{aligned}
    = \mathbb{E}_{\mathbf{x} \sim p_{data}} \left[\log \left( \sum_\mathbf{z_1} \sum_\mathbf{z_2} q_{\phi}(\mathbf{z_1|x}).q_{\phi}(\mathbf{z_2|x})  \right. \right. \\ \times
    \left. \left.  \frac{p_\theta(\mathbf{z_1}). p_\theta(\mathbf{z_2})}{q_{\phi}(\mathbf{z_1|x}). q_{\phi}(\mathbf{z_2|x})} . p_\theta(\mathbf{x|z_1, z_2}) \right) \right]
    \end{aligned}
    \label{intro_aux}
\end{equation}
Simplifying,
\begin{equation}
    \begin{aligned}
    (\ref{intro_aux}) = \mathbb{E}_{\mathbf{x} \sim p_{data}} \left[ \log \left(\sum_{\mathbf{z_1} \sim p_\theta(\mathbf{z_1})} \frac{p(\mathbf{z_1})}{q_\phi(\mathbf{z_1 | x)}} \right. \right. \\ \left. \left. \times
    \sum_{\mathbf{z}_2 \sim p_{\theta}(\mathbf{z_2})} \frac{p(\mathbf{z_2})}{q_\phi(\mathbf{z_2 | x)}} . p_\theta(\mathbf{x|z_1, z_2}). q_{\phi}(\mathbf{x|z_1, z_2}) \right) \right]
    \end{aligned}
    \label{simplifying}
\end{equation}
Applying Jensen's inequality, 


\begin{equation}
    \begin{aligned}
    (\ref{simplifying}) \ge \mathbb{E}_{\mathbf{x} \sim p_{data}} \left[ \mathbb{E}_{\mathbf{z_1} \sim q_\phi(\mathbf{z_1|x)}} \log \left( \frac{p_\theta(\mathbf{z_1})}{q_{\phi}(\mathbf{z_1|x})} \right)  \right. \\ \left. +
     \mathbb{E}_{\mathbf{z_2} \sim q(\mathbf{z_2|x)}} \log \left( \frac{p_\theta(\mathbf{z_2})}{q_{\phi}(\mathbf{z_2|x})} \right) \right. \\ \left. + \mathbb{E}_{q_\phi(\mathbf{z_1, z_2|x})} \log(p_\theta (\mathbf{x|z_1, z_2})) \right]
    \end{aligned}
    \label{jensen}
\end{equation}

Now we can write each term in their $KL$ equivalence. 

\begin{equation}
    \begin{aligned}
    (\ref{jensen}) = \mathbb{E}_{\mathbf{x} \sim p_{data}} \left[-KL(q_\phi(\mathbf{z_{1}|x})||p_\theta(\mathbf{z_1})) \right. \\ \left. -  KL(q_\phi(\mathbf{z_{2}|x})||p_\theta(\mathbf{z_2})) \right.  \\ + \left.
    \mathbb{E}_{q_{(\mathbf{z_1, z_2|x}})} \log(p_\theta(\mathbf{x|z_1, z_2})) \right]
    \end{aligned}
    \label{kl_equiv}
\end{equation}

We can re-arrange equation \ref{kl_equiv} in entropy terms:
\begin{equation}
    \begin{aligned}
    = \mathbb{E}_{\mathbf{x} \sim p_{\mathbf{data}}} \left[ \mathbb{E}_{q_\phi(\mathbf{z_1, z_2|x})} \log \left( p_\theta(\mathbf{x|z_1, z_2}) \right) \right. \\ \left. + H(q_\phi(\mathbf{z_1|x})) + H(q_\phi(\mathbf{z_2|x}))  \right. \\ \left. +  \mathbb{E}_{q_\phi(\mathbf{z_1})} \log \left(p_\theta(\mathbf{z_1})) \right. \right. \\ 
    \left. + \mathbb{E}_{q_\phi(\mathbf{z_2})} \log \left(p_\theta(\mathbf{z_2})
    \right)\right]\\
    \end{aligned}
\end{equation}
Further re-arranging, we arrive at the same form as equation \ref{entropy}:
\begin{equation}
    \begin{aligned}
    = \mathbb{E}_{\mathbf{x} \sim p_{\mathbf{data}}} \left[ \mathbb{E}_{q_\phi(\mathbf{z_1, z_2|x})} \log \left( p_\theta(\mathbf{x|z_1, z_2}) \right) \right.  \\ \left. + H(q_\phi(\mathbf{z_1|x})) + H(q_\phi(\mathbf{z_2|x})) \right. \\ \left. - H \left(p_\theta(\mathbf{z_1})\right)  -  H \left(p_{\theta}(\mathbf{z_2})
    \right)\right].\\
    \end{aligned}
    \label{proof}
\end{equation}

This proves equation (\ref{entropy}).

\bibliographystyle{seg}  
\bibliography{ref.bib}
\end{document}